%% file: main.tex
\documentclass[sigconf,nonacm]{acmart}
\input{setup.tex}
\newif\ifblind
\blindfalse

\begin{document}
\title{SPEChpc 2021 Benchmarks on Ice Lake and Sapphire Rapids Infiniband Clusters: A Performance and Energy Case Study}
\ifblind
\else
\author{Ayesha Afzal}
\email{ayesha.afzal@fau.de}
\orcid{0000-0001-5061-0438}
\affiliation{%
	\institution{Erlangen National High Performance Computing Center (NHR@FAU)}
	\city{91058 Erlangen}
	\country{Germany}}

\author{Georg Hager}
\email{georg.hager@fau.de}
\orcid{0000-0002-8723-2781}
\affiliation{%
	\institution{Erlangen National High Performance Computing Center (NHR@FAU)}
	\city{91058 Erlangen}
	\country{Germany}}

\author{Gerhard Wellein}
\email{gerhard.wellein@fau.de}
\orcid{0000-0001-7371-3026}
\affiliation{%
	\institution{Department of Computer Science, Friedrich-Alexander-Universität}
	\city{Erlangen-N\"urnberg}
	\country{Germany}
	\postcode{91058}}

\renewcommand{\shortauthors}{Afzal et al.}
\fi
\pagestyle{plain}

\begin{abstract}
In this work, fundamental performance, power, and energy characteristics of the full SPEChpc 2021 benchmark suite are assessed on two different clusters based on Intel Ice Lake and Sapphire Rapids CPUs using the MPI-only codes' variants. We use memory bandwidth, data volume, and scalability metrics in order to categorize the benchmarks and pinpoint relevant performance and scalability bottlenecks on the node and cluster levels. Common patterns such as memory bandwidth limitation, dominating communication and synchronization overhead, MPI serialization, superlinear scaling, and alignment issues could be identified, in isolation or in combination, showing that SPEChpc 2021 is representative of many HPC workloads. Power dissipation and energy measurements indicate that the modern Intel server CPUs have such a high idle power level that race-to-idle is the paramount strategy for energy to solution and energy-delay product minimization. On the chip level, only memory-bound code shows a clear advantage of Sapphire Rapids compared to Ice Lake in terms of energy to solution.
\end{abstract}


\maketitle

\section{Introduction and related work}
Modern HPC systems and programming models are becoming more complicated, heterogeneous, and diversified, making it difficult to evaluate performance and aim for performance portability. 
This necessitates a carefully crafted benchmark collection to design the hardware and software stacks of future large-scale systems.

Numerous benchmark suites and mini-appli\-cations have been developed in the HPC field.
The most popular ones are the
HPC Challenge (HPCC) benchmark suite~\cite{Luszczek2006},
the NAS parallel benchmarks (NPB)~\cite{Bailey91},
the Standard Performance Evaluation Cor\-po\-ration/High-Per\-for\-mance Group (SPEC/HPG) first SPEC HPC96 benchmark suite~\cite{Eigenmann1996} improved by SPEC HPC2002~\cite{Eigenmann2002}, 
the Scalable Heterogeneous Computing (SHOC) benchmark suite~\cite{Danalis2010}, 
LINPACK~\cite{dongarra2003},
the HPCG benchmark~\cite{Dongarra2016},
HPGMG~\cite{Adams2014} for multigrid methods,
the HPL-AI benchmark~\cite{Kudo2020} and HPC-MixPBench for mixed-precision analysis~\cite{Parasyris2020}, 
HPC AI500 for HPC AI systems~\cite{Jiang2018} and 
the GPCNeT benchmark suite for contention analysis in HPC networks~\cite{Chunduri2019}. 
A few more examples of mini-applications are MiniFE, MiniMD, phdMesh, MiniXyce, Prolego, and others~\cite{Crozier2009}.

Currently, the {HPG} is actively maintaining four benchmark suites: \CODE{SPEC MPI 2007} (MPI programming model)~\cite{Muller2010}, \CODE{SPEC OMP 2012} (OpenMP programming model)~\cite{Muller2012}, \CODE{SPEC Accel} (X programming model, where X can be OpenCL, OpenACC, and OpenMP target offload for accelerators)~\cite{Juckeland2015}, and \CODE{SPEChpc 2021}\footnote{
\textsc{SPEC} benchmark suites:
\textsc{SPEC} MPI (\textsc{TM}) 2007 <{\url{https://spec.org/mpi2007}}>,
\textsc{SPEC} OMP (\textsc{TM}) 2012 <{\url{https://spec.org/omp2012}}>, 
\textsc{SPEC} ACCEL (\textsc{TM}) <{\url{https://spec.org/accel}}>
\textsc{SPEC}hpc (\textsc{TM}) 2021 <{\url{https://spec.org/hpc2021}}>
} (MPI and hybrid MPI+X programming models for both CPU-only and heterogeneous {HPC} systems with multiple accelerators)~\cite{Li2022,Brunst2022}.

\mypara{Contributions}
This work's primary contributions are as follows: 
    We provide an overview of the full SPEChpc 2021 benchmark suite in MPI-only mode and provide performance and energy metrics on ccNUMA domain, node, and multi-node levels on two clusters with different generations of modern Intel server CPUs.
    We further pinpoint scalability and performance issues and identify their root causes, demonstrating the value of fundamental resource metrics like data volume and bandwidths. 
    Finally, we show that analyzing power dissipation and energy consumption requires a clear distinction between memory-bound and non-memory-bound codes and that the minimization of energy-delay product and energy are dominated by chip idle power and code scaling characteristics.

\mypara{Overview}
This paper is organized as follows:
We introduce the SPEChpc 2021 Benchmarks in Sect.~\ref{sec:SPECintro} and describe our experimental setup and methodology in Sect.~\ref{sec:setup}.
We then discuss SPEChpc 2021 parallel benchmark results:
In Sect.~\ref{sec:NL} we focus on node-level performance, power, and energy using the \emph{tiny} workloads.
Similarly, Sect.~\ref{sec:MNL} uses the \emph{small} workloads for multi-node analysis.
Finally, Sect.~\ref{sec:conclusion} summarizes the paper and gives an outlook to future work.

\section{SPEChpc 2021 Benchmarks}
\label{sec:SPECintro}

The SPEChpc 2021 collection, released in October 2021, covers a wide spectrum of science and engineering programs that are representative of {HPC} workloads and are portable across CPUs and accelerators. 
It aims to be the industry standard for assessing the efficiency of parallel computing workloads on heterogeneous platforms.
The benchmarks are accessible on the SPEC website for non-profit use.

\mypara{Implementation}
\input{figures/SPEChpc2021}
Table~\ref{tab:SPEChpc2021} lists the names of all nine benchmarks, their input configurations for tiny and small workloads, their programming language, the lines of code, and the employed collective communication primitives. 
Table~\ref{tab:SPECbrief} briefly outlines the numerical information and the prospective application domain of benchmarks.
The SPEChpc 2021 benchmarks use multiple programming languages (Fortran, C, and C++) and parallel programming models (MPI, MPI+OpenACC, MPI+OpenMP, and MPI+OpenMP with target off{}load) and are integrated with a benchmarking harness to ensure results correctness and sensible reporting.

\mypara{Workload suites}
To meet the need for different system sizes, the four suites \enquote{tiny}, \enquote{small}, \enquote{medium} and \enquote{large} allow running the benchmark suite  from one to hundreds of nodes. 
The {SPEC} benchmarks concentrate on compute-intensive parallel performance. Each benchmark distributes the same workload over any number of active processes or threads (strong scaling).
The {\{tiny, small, medium, large\} workloads} utilize up to {$\{$0.06, 0.48, 4, 14.5$\}$~\TB} of memory and are, according to the documentation, designed to run on clusters using {\{1--256, 64--1024, 256--4096, 2048--32768\} processes}, respectively.

\mypara{SPEChpc 2021 benchmark set-up}
This work focuses on the MPI versions of the SPEChpc benchmarks on CPU-only systems.
Our goal is not to achieve best performance by choosing the appropriate programming model but rather to pinpoint peculiarities of the various benchmark codes. MPI+X hybrid-parallel will surely have their own set of issues and warrant their own investigation.
Furthermore we expect some insights from MPI-only variants to be relevant on larger scales and hybrid programming models, such as the prime number problem or the cutting problem; see Sections~\ref{sec:NL} and \ref{sec:MNL}.

We only include findings for the \emph{tiny} and \emph{small} workloads since the \emph{medium} and \emph{large} workloads are only supported by six out of the nine benchmarks. 
Single-node performance is examined first, using the \emph{tiny} workloads; after that we turn to multi-node performance using the \emph{small} workloads and up to 1664 MPI processes on both clusters.
For additional information we refer to our detailed performance data artifact appendix\footnote{\url{https://doi.org/10.5281/zenodo.8338037}\label{foot:AD}}. 

\section{Hardware-software setup} 
\label{sec:setup}
The hardware and software environments employed for all experiments are shown in Table~\ref{tab:systems}.
Two Intel-based InfiniBand (HDR-100) clusters were at our disposal:
\begin{enumerate}
    \item ClusterA\footnote{\ifblind
\else\url{https://hpc.fau.de/systems-services/documentation-instructions/clusters/fritz-cluster}\label{foot:Fritz}\fi} comprising two Intel Xeon Ice Lake CPUs per node with 36 cores each
    \item ClusterB\ifblind
\else\footref{foot:Fritz}\fi comprising two Intel Xeon Sapphire Rapids CPUs per node with 52 cores each
\end{enumerate}
Hyper-threading was disabled on both systems. Consecutive MPI processes were mapped to consecutive cores using the \CODE{likwid-mpirun}~\cite{LIKWID} startup wrapper.  
Sub-NUMA Clustering (SNC) was activated on both systems, leading to a fundamental scaling unit (i.e., one ccNUMA domain) of half (i.e., 18 cores) and one-fourth (i.e., 13 cores) of a socket on ClusterA and ClusterB, respectively.
All prefetching mechanisms in the hardware were enabled.
We always employed the widest SIMD instruction set, i.e., \CODE{AVX-512}, supported by the Intel architectures.
The clock frequency of the ClusterA and ClusterB nodes was consistently fixed to the base values of their respective CPUs via the SLURM batch scheduler (option \CODE{--cpu-freq}).
The expected clock frequency was verified with the \CODE{likwid-perfctr} utility, which was also used for reading hardware performance events. 
On ClusterB, we had to employ a beta version of the LIKWID suite that supports Sapphire Rapids CPUs. 
This work used the first official release (version 1.0.3) of the SPEChpc 2021 suite.

\input{figures/SPEC_brief}
\input{figures/tab_systems.tex}
\begin{figure*}[t]
    \begin{minipage}[c]{0.76\textwidth}
    \begin{minipage}{1.15\textwidth}
        \subfloat[ClusterA speedup]
        {\includegraphics[scale=0.32]{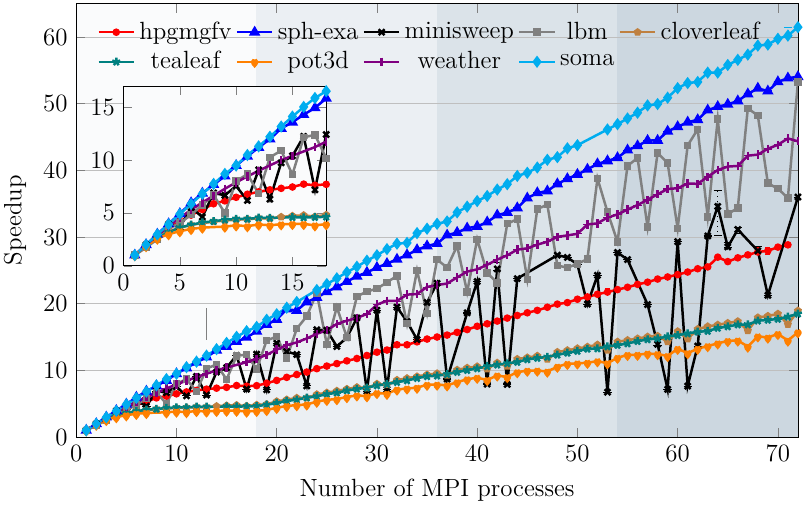}}\hspace{0.1em}
        \subfloat[ClusterA performance] 
        {\includegraphics[scale=0.32]{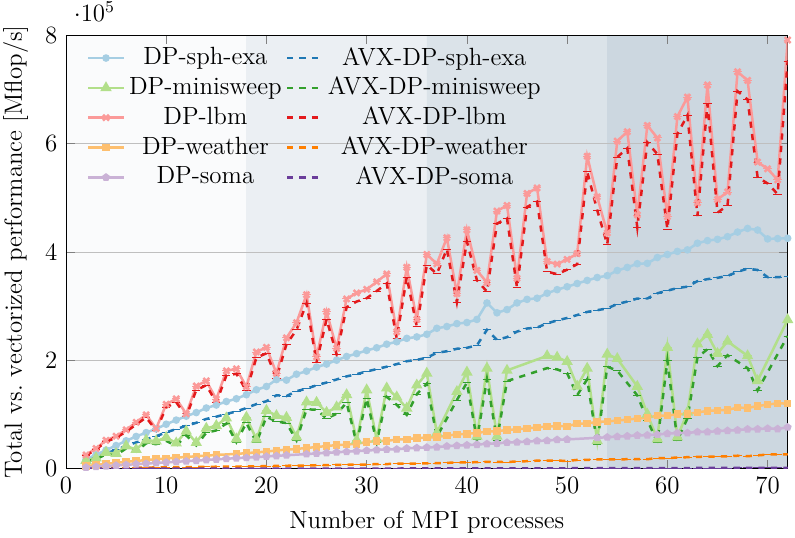}}\hspace{0.1em}
        \subfloat[ClusterA performance]
        {\includegraphics[scale=0.32]{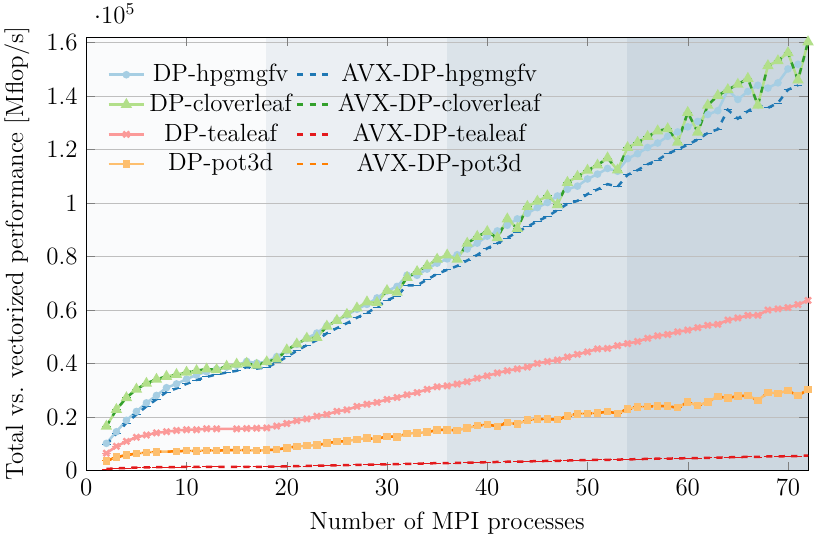}}
        \label{fig:NLSClusterA}
    \end{minipage}%
    
    \begin{minipage}{1.15\textwidth}
        \subfloat[ClusterB speedup \\(all nine codes)]
        {\includegraphics[scale=0.32]{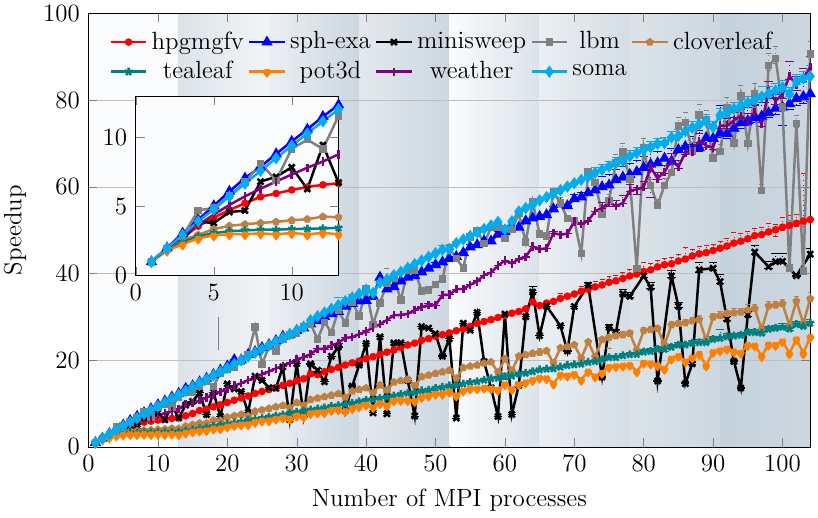}}\hspace{0.1em}
        \subfloat[ClusterB performance (non-memory-bound codes)]{\includegraphics[scale=0.32]{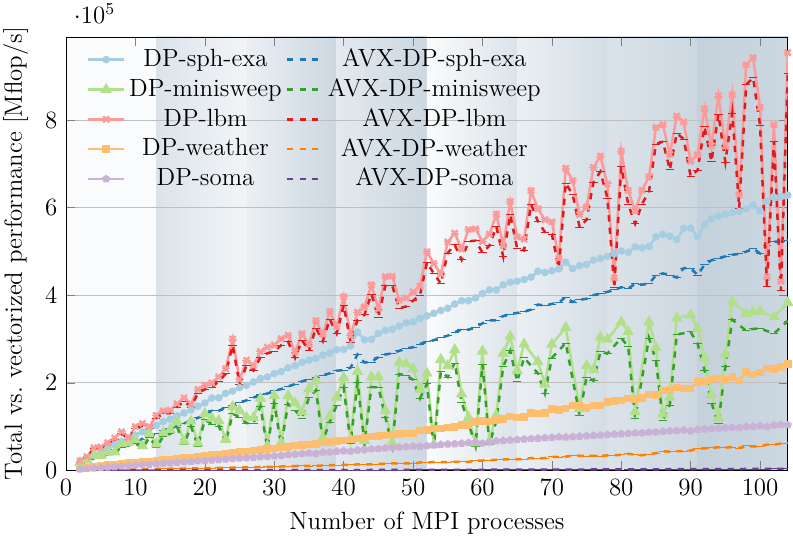}}\hspace{0.1em}
        \subfloat[ClusterB performance \\(memory-bound codes)]{\includegraphics[scale=0.32]{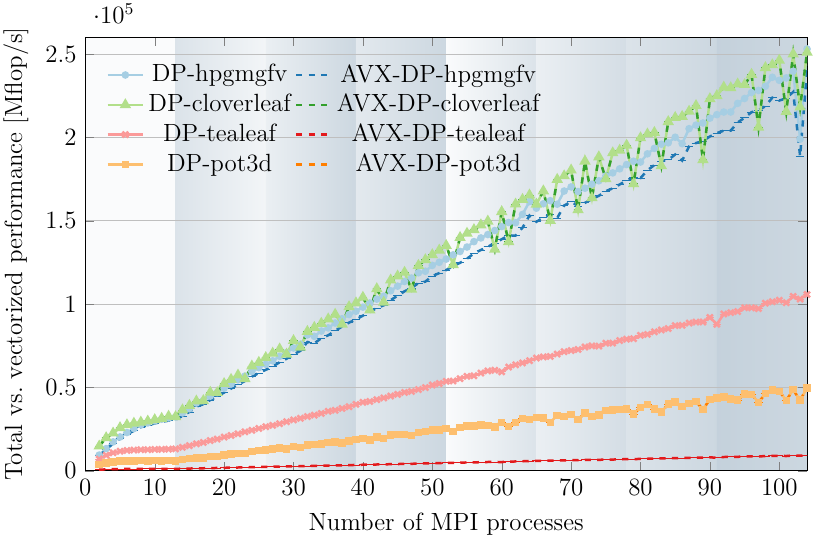}}
        \label{fig:NLSClusterB}
    \end{minipage}
    \end{minipage}\hfill
    \begin{minipage}[c]{0.24\textwidth}
    \caption{
    SPEChpc 2021 tiny suite performance on a node of ClusterA (top) and ClusterB (bottom).
    The shaded background marks the ccNUMA domains of each cluster.
    Two codes (\CODE{lbm} and \CODE{minisweep}) exhibit intriguing patterns that hold up across multiple runs on each system.
    (a, d) The speedup (min, max, average) on the first ccNUMA domain is shown in an inset.
    (b-c, e-f) 
    A well-vectorized code has a small difference between \CODE{DP} (actual performance) and \CODE{DP-AVX} (vectorized part only). 
    }
    \label{fig:singleNodeSpeedupDP}
    \end{minipage}
\end{figure*}
Time-resolved Roof{}line plots\footref{foot:AD} of the benchmarks were obtained using the \CODE{ClusterCockpit} monitoring framework~\cite{Clustercockpit:2019}.
The Intel Trace Analyzer and Collector (\CODE{ITAC}) tool\footnote{The \CODE{ITAC} utility synchronizes clocks among MPI processes:
\url{https://intel.com/content/www/us/en/develop/documentation/itac-user-and-reference-guide/top/intel-trace-collector-reference/time-stamping/clock-synchronization.html}} was used for visualizing MPI event traces.
Since RAPL measurements vary across nodes, all benchmarks were run on the same node for the node-level analysis of Sect.~\ref{sec:NL}. 
The working sets of the \emph{tiny} or \emph{small} suites were at least ten times the size of the \LLC of one node, which prevented it from fitting into the available cache\footnote{The \LLC (LLC) is made up of the non-inclusive victim L3 and the L2 caches in the Ice lake and Sapphire Rapids processors.}. Nevertheless, cache effects could be observed in multi-node scaling for some of the codes. 
Memory bandwidths were determined using the ratio of memory data volume to wall-clock time.
Before performing the measurements, at least two warm-up time steps, including global synchronisation, were conducted to allow the MPI runtime to stabilize and eliminate first-call overhead.
To account for variations in runtime, we repeated code executions several times and only statistically significant deviations were reported.

\begin{figure*}[t]
    \centering
    \subfloat[ClusterA memory bandwidth]{\includegraphics[scale=0.34]{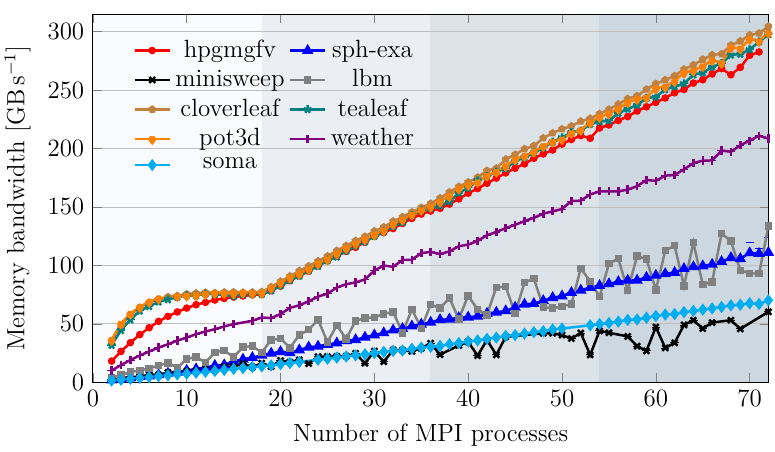}}
    \subfloat[ClusterB memory bandwidth]{\includegraphics[scale=0.34]{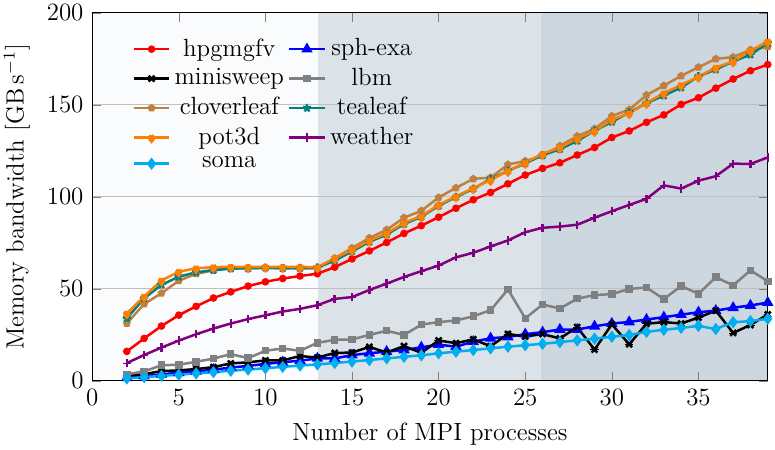}}
    \subfloat[L3 bandwidth]{\includegraphics[scale=0.33]{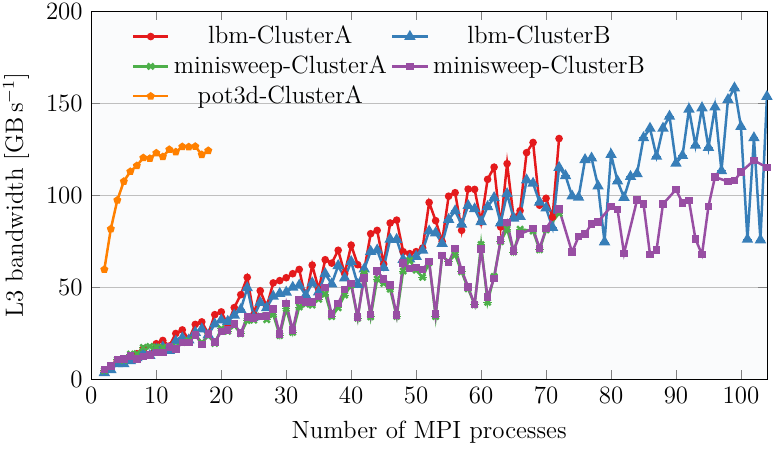}}
    \subfloat[L2 bandwidth]{\includegraphics[scale=0.33]{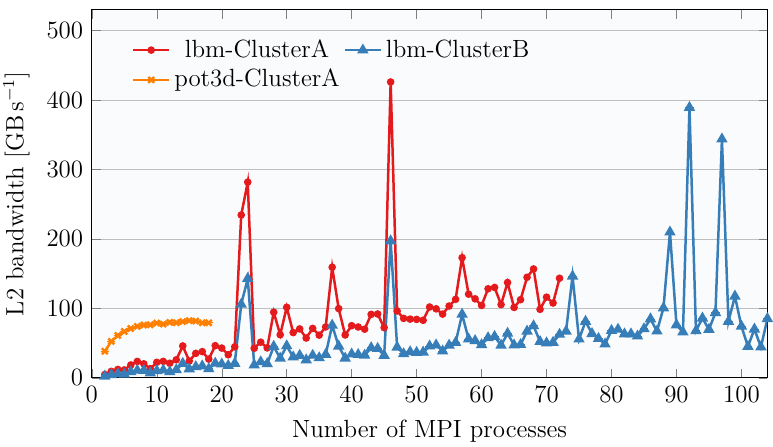}}\quad    
    \subfloat[ClusterA memory data volume]{\includegraphics[scale=0.34]{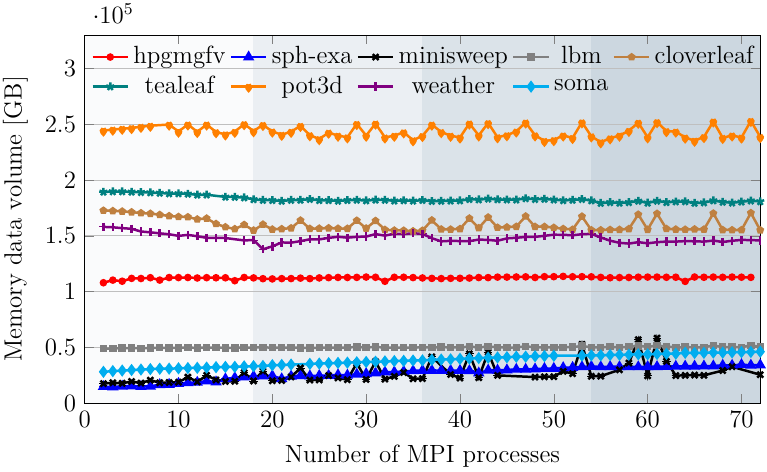}} 
    \subfloat[ClusterB memory data volume]{\includegraphics[scale=0.34]{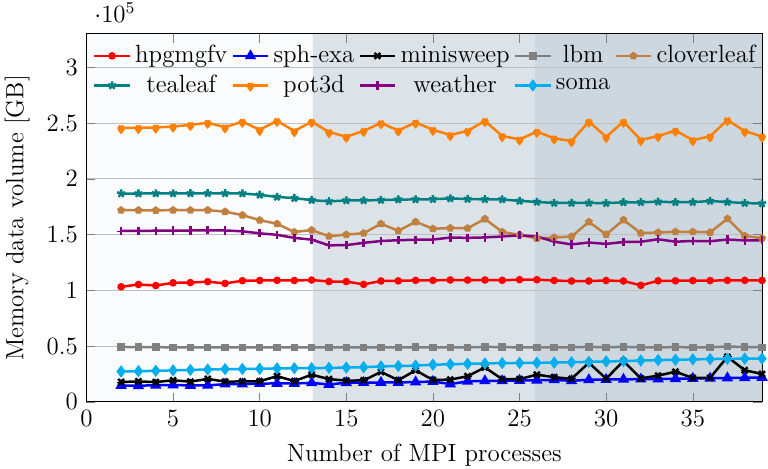}} 
    \subfloat[L3 data volume]{\includegraphics[scale=0.34]{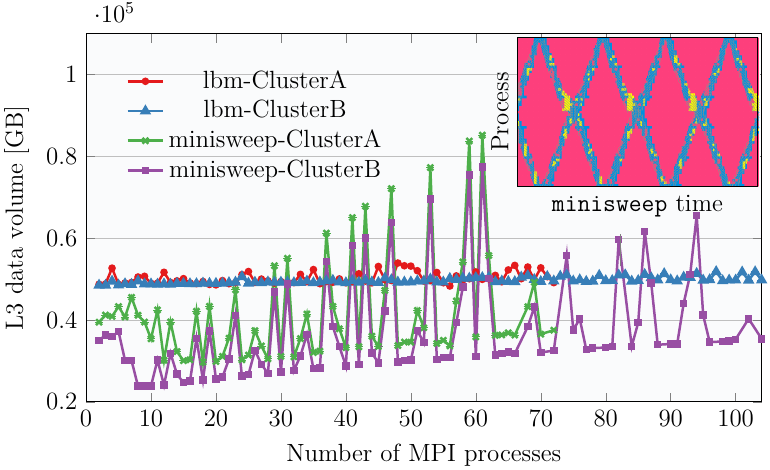}}
    \subfloat[L2 data volume]{\includegraphics[scale=0.34]{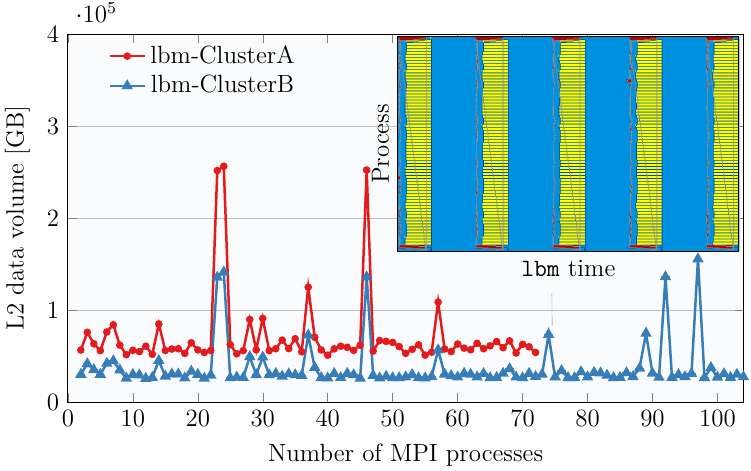}}        
    \caption{Node-level bandwidth and data volume behavior of (a-b, e-f) memory, (c, g) L3 cache, and (d, h) L2 cache for the SPEChpc 2021 tiny suite on both clusters. The background shading layers denote the ccNUMA domains.
    On ClusterA, timeline inset displays in (g) the \CODE{minisweep} time spent in \CODE{MPI\_Recv} (red), \CODE{computation} (blue), and  \CODE{MPI\_Send} (yellow) for 59 processes, while in (h) the \CODE{lbm} time spent in \CODE{MPI\_Wait} (red), \CODE{computation} (blue), and  \CODE{MPI\_Barrier} (yellow) for 71 processes.
    }
    \label{fig:singleNodeMEMCaches}
\end{figure*}
\section{Node-level analysis}
\label{sec:NL}
The ``tiny'' workload suite was used for node-level analysis. 
We first examine each code's scalability, memory-boundness, vectorization, and the underlying causes of scaling issues. We then determine the impact of these findings on power and energy consumed by ``hot'' and ``cold'' codes and the relevance of the baseline (idle) power and energy-delay product. 

\subsection{Performance and speedup}
\label{sec:NLP}
In this section we show  how the scalability, vectorization, process timeline, memory bandwidth, and data transfer volume can be used to find the underlying causes of non-ideal scalability.

\subsubsection{Speedup}
A saturation pattern, i.e., the speedup approaching a limit across the cores of a ccNUMA domain, is an indicator for memory-bound behavior.
Lacking other bottlenecks, the speedup \emph{across} ccNUMA domains should be ideal, i.e., a factor of 4 (ClusterA) or 8 (ClusterB)
unless cache effects allow for superlinear scaling.
With a baseline of ccNUMA domain, we can extract  the following parallel efficiencies (in percent) from Figure~\ref{fig:singleNodeSpeedupDP}(a, d):
\vspace{-1em}
\begin{table}[h!]
	\begin{adjustbox}{width=0.48\textwidth}
    \Huge
    \pgfplotstabletypeset[
    every column/.style=,
    color cells={min=100,max=130},
    /pgf/number format/fixed,
    /pgf/number format/precision=3,
    col sep=comma,
    every head row/.style={%
    	before row={\hline
    	},
    	after row=\hline
    },
    ]{
    	Speedup percentage,lbm, soma, tealeaf, cloverleaf, minisweep, pot3d, sph-exa, hpgmgfv, weather
        ClusterA,130, 93, 100, 98, 73, 100, 80, 95, 95 
        ClusterB,95, 86, 100, 96, 80, 104, 79, 98, 121
    }
    \end{adjustbox}
\end{table}
\vspace{-1em}

On both clusters, \CODE{lbm} and \CODE{minisweep} show reproducible fluctuations in scalability which makes the speedup across domains less meaningful. 
The superlinear scaling for \CODE{weather} on ClusterB is caused by cache effects as the Sapphire Rapids CPU has significantly more aggregate outer-level cache; this shows the fact that the non-memory-bound \CODE{weather} benchmark still contains some memory-intensive kernels.

\subsubsection{Performance}
Adopting a Roof{}line-like view of hardware-software interaction, performance metrics allow to compare application performance differences with hardware properties like peak performance and memory bandwidth. 
According to Table~\ref{tab:systems}, comparing ClusterB with ClusterA the ratio of peak performance and memory bandwidth is 1.2 and 1.5  respectively.
We therefore expect a node of ClusterB to be 1.2 to 1.5 times faster than a node of ClusterA, depending on whether the code is compute bound or memory bound. 
From Figure~\ref{fig:singleNodeSpeedupDP}(b-c, e-f) we can read the following actual performance ratios:
\vspace{-1em}
\begin{table}[h!]
	\begin{adjustbox}{width=0.48\textwidth}
    \Huge
    \pgfplotstabletypeset[
    every column/.style=,
    color cells={min=1.5,max=1.55},
    /pgf/number format/fixed,
    /pgf/number format/precision=3,
    col sep=comma,
    every head row/.style={%
    	before row={\hline
    		&\multicolumn{4}{c}{Non-memory-bound codes} & \multicolumn{5}{c}{Memory-bound codes}\\
    	},
    	after row=\hline
    },
    ]{
    	Acceleration factor,lbm, soma, sweep, sph-exa, weather, tealeaf, cloverleaf, pot3d, hpgmgfv
        ClusterB over ClusterA,1.21, 1.35, 1.39, 1.48, 2.03, 1.66, 1.57, 1.63, 1.65 
    }
    \end{adjustbox}
\end{table}
\vspace{-1em}

In a number of applications, the speedup of Sapphire Rapids over Ice Lake exceeds the expected ratio. This can be attributed to architectural enhancements introduced in Sapphire Rapids, such as the larger L2 and L3 caches and L3 bandwidth\footnote{ClusterB' core comprises 45\% more L3 cache and 60\% more L2 cache than ClusterA.}.
This applies to both memory-bound and non-memory-bound codes, as the latter can still be cache sensitive.


\subsubsection{Vectorization}
The vectorization (SIMD) analysis on both systems is shown in Figures~\ref{fig:singleNodeSpeedupDP}(b-c) and (e-f) separately for memory-bound and non-memory-bound codes.
The chosen compiler flags allow the compiler to utilize AVX-512 instructions, and all nine benchmarks primarily use them on both CPUs.
From the data we can extract the following vectorization ratios, which we define as the ratio of actual numerical work (flops) done with SIMD instructions to the overall numerical work:

\begin{table}[h!]
	\begin{adjustbox}{width=0.48\textwidth}
    \Huge
    \pgfplotstabletypeset[
    every column/.style=,
    color cells={min=70,max=75},
    /pgf/number format/fixed,
    /pgf/number format/precision=3,
    col sep=comma,
    every head row/.style={%
    	before row={\hline
    	},
    	after row=\hline
    },
    ]{
    	Vectorization percentage,lbm, soma, tealeaf, cloverleaf, minisweep, pot3d, sph-exa, hpgmgfv, weather
        ClusterA/ ClusterB,95.1, 2.2, 8.8, 100, 89.1, 99.9, 83.3, 94.8, 22.2 
    }
    \end{adjustbox}
\end{table}
\vspace{-0.8em}

The percentage of vectorized work is similar on both systems.
The memory-bound \CODE{cloverleaf} and \CODE{pot3d} codes and the most compute-intensive \CODE{lbm} show the highest vectorization ratio.
However, the memory-bound \CODE{tealeaf} and the non-memory-bound \CODE{soma} code are poorly vectorized. 
Looking at the memory bandwidth of \CODE{weather} (see next section), it is probable that it might become fully memory bound if it could be efficiently vectorized.

\subsubsection{Bandwidth}
Figure~\ref{fig:singleNodeMEMCaches}(a-b) presents node memory bandwidth measurements of all benchmarks on ClusterA and ClusterB.
Five of the nine benchmarks (\CODE{hpgmgfv, clover\-leaf, tea\-leaf, pot3d, weather}) draw a significant fraction of the available memory bandwidth of the node, with only the first four actually achieving the saturated memory bandwidth on a ccNUMA domain (75--78~\GBS\ for ClusterA and 58--62~\GBS\ for ClusterB).
Among these four, the strongly saturated \CODE{pot3d}, \CODE{cloverleaf}, and \CODE{tealeaf} show strong saturation patterns, whereas \CODE{hpgmgfv} is only weakly saturating and becomes less memory-bound with more cores.
\CODE{weather}'s memory bandwidth behavior on a ccNUMA domain indicates the presence of memory-bound and non-memory-bound kernels, with the latter dominating the runtime on both systems.
In memory-bound codes, the fact that L3 cache on ClusterA has a greater bandwidth than L2 (124~\GBS\ vs.\ 80~\GBS\ for \CODE{pot3d}) indicates that L3 is a victim cache (with memory prefetchers enabled) and sees additional traffic coming down from L2; see orange data points in Figure~\ref{fig:singleNodeMEMCaches}(c-d).

\begin{figure*}[t]
    \centering
    \begin{minipage}{0.49\textwidth}
    \hspace{-0.9em}
        \centering
        {\includegraphics[scale=0.32]{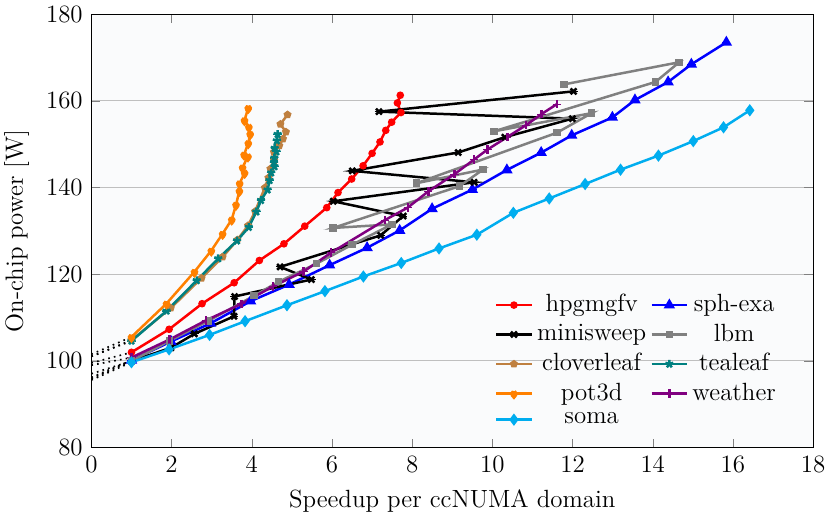}}{\includegraphics[scale=0.32]{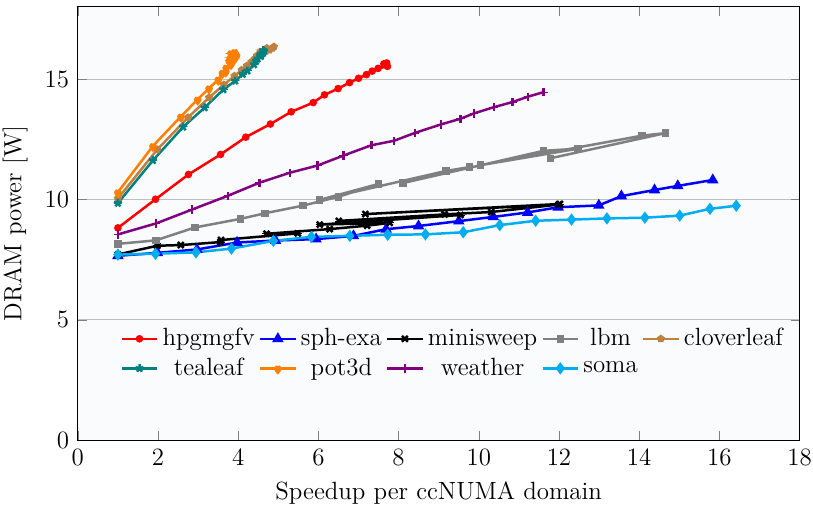}}
        \caption*{(a) ClusterA ccNUMA domain power}
        \label{fig:DLPClusterA}
    \end{minipage}%
    \begin{minipage}{0.49\textwidth}
        \centering
        {\includegraphics[scale=0.32]{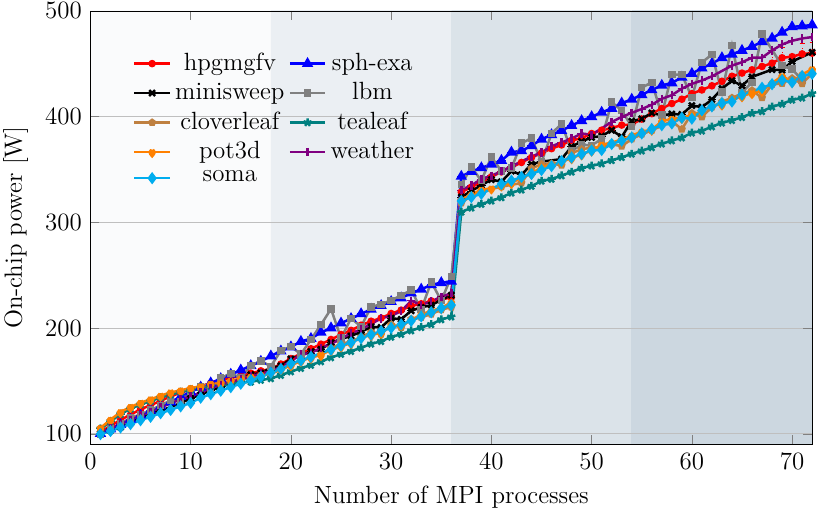}}{\includegraphics[scale=0.32]{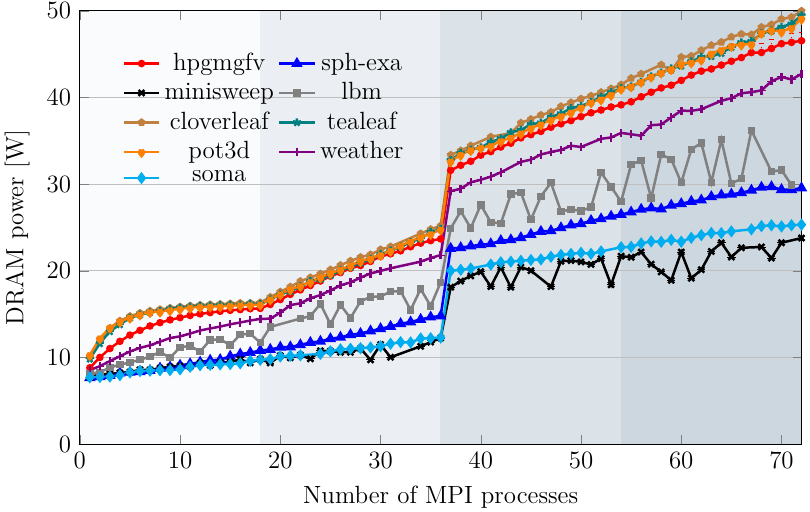}}
        \caption*{(b) ClusterA node power}
        \label{fig:NLPClusterA}
    \end{minipage}%

    \begin{minipage}{0.49\textwidth}
    \hspace{-0.7em}
        {\includegraphics[scale=0.32]{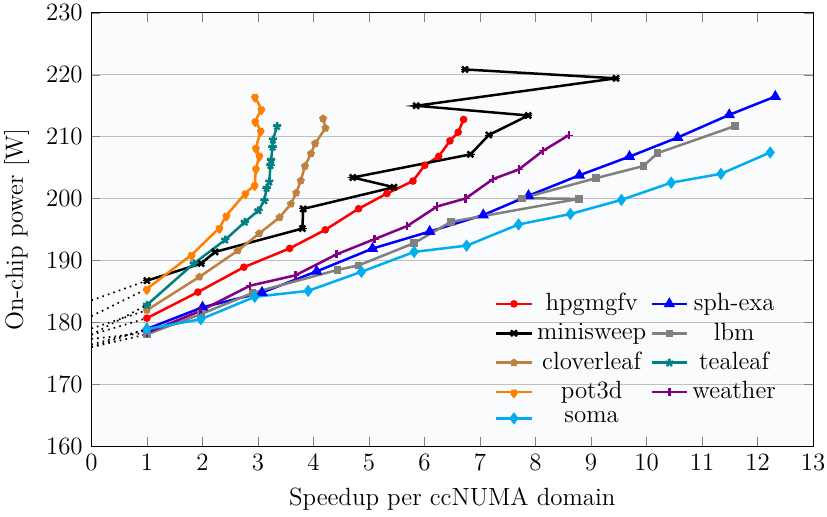}}\hspace{-0.2em}
        {\includegraphics[scale=0.32]{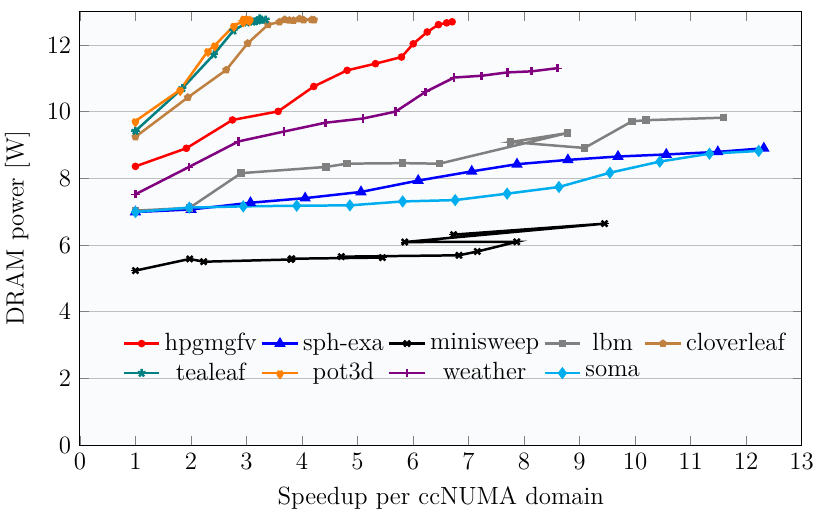}}
        \caption*{(c) ClusterB ccNUMA domain power}
        \label{fig:DLPClusterB}
    \end{minipage}%
    \begin{minipage}{0.49\textwidth}
        {\includegraphics[scale=0.32]{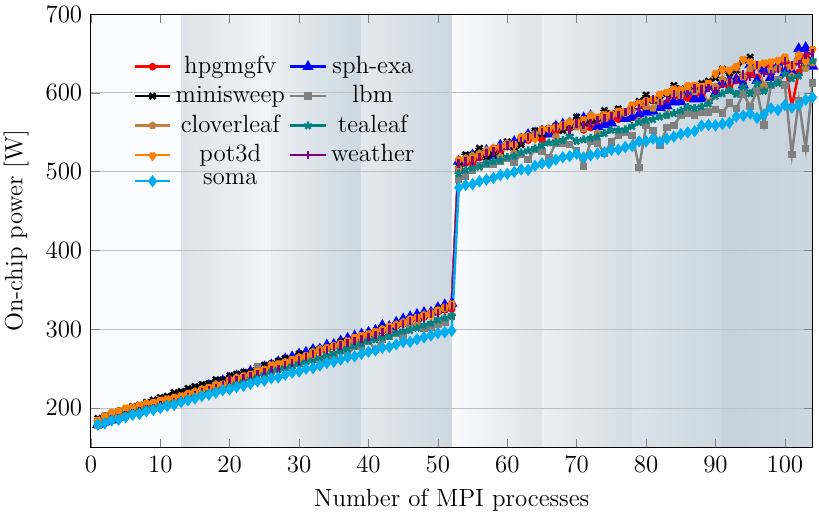}}
        {\includegraphics[scale=0.32]
        {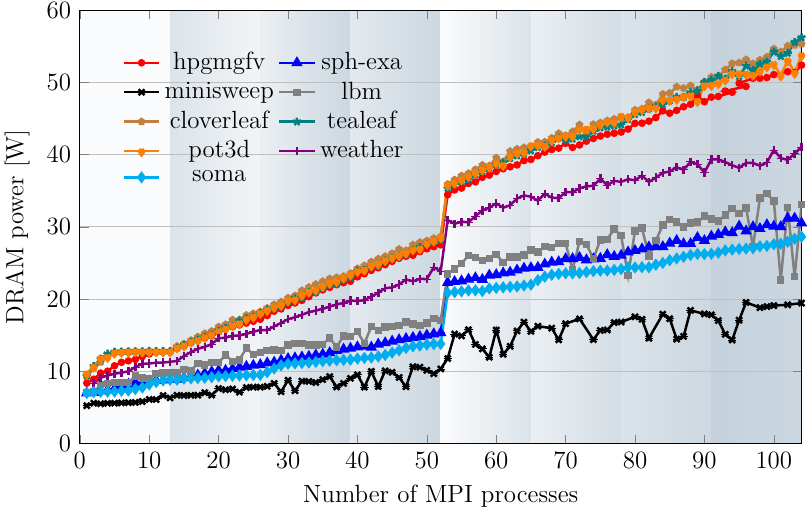}}
        \caption*{(d) ClusterB node power}
        \label{fig:NLPClusterB}
    \end{minipage}
    \caption{
    SPEChpc 2021 tiny suite power dissipation of CPUs and DRAM (via RAPL); (a, c) on one ccNUMA domain, showing power vs.\ speedup, and (b, d) full intra-node node for both clusters.
    In (a, c), the baseline power was determined by extrapolating o zero cores (black dotted lines).
    In (b, d) the ccNUMA domains are indicated by the background shading layers.
    }
    \label{fig:singleNodePower}
\end{figure*}
\subsubsection{Consistent fluctuations in \CODE{minisweep}}
The \CODE{minisweep} code shows consistent performance fluctuations with changing core count. Specifically, prime numbers and some other special numbers of processes such as \{9, 26, 34, 51, 69\}  are detrimental for performance; see Figure~\ref{fig:singleNodeMEMCaches}.
For example, on ClusterA, performance drops by 75\% from 58 to 59 processes, where 75\% of the time is spent in \CODE{MPI\_Recv}, 5.5\% in \CODE{MPI\_Sendecv}, and 19.5\% in computation.
The root cause is a communication serialization performance bug as shown by the ITAC timeline of 59 processes in the inset of Figure~\ref{fig:singleNodeMEMCaches}(g).
The \CODE{minisweep} code uses open boundary conditions and synchronous rendezvous mode (due to large messages). 
Traces show that every process sends to its top neighbor first; with open boundary conditions, only the top process in the chain does not have a top neighbor and can thus call \CODE{MPI\_Recv} right away. Subsequently, the communication ``ripples'' through the processes, leading to massive MPI waiting times.

\subsubsection{Consistent fluctuations in \CODE{lbm}}

The non-memory-bound \CODE{lbm} code includes
a strongly memory-bound ``propagate'' kernel performing sparse memory accesses and a ``collide'' kernel 
with about 6600 floating-point operations per lattice site update and consequently high intensity.
In contrast to \CODE{minisweep}, the performance scaling for \CODE{lbm} (see Figure~\ref{fig:singleNodeSpeedupDP}) shows large fluctuations with clear upper and lower limits. In-cache effects can not be observed in memory or L3 data volume; see Figure~\ref{fig:singleNodeMEMCaches}(e-h).
The drop in performance is not always accompanied by excess L2 data volume, indicating several overlapping effects.
For instance, L2 data traffic peaks at \{22, 23, 31 and 45\} processes but poor performance is observed also with \{47--50, 68--71\} processes. 
Powers of two in both the x and y directions, such as 4096 and 16384, are particularly susceptible to these fluctuations.
The 44 and 45 processes are not that dissimilar from a domain decomposition and layer condition point of view, but they differ greatly in terms of L2 data transfer and traffic.
Further, in addition to drawing huge L2 data volume, runs at 22 and 23 processes also consume more L3 and memory bandwidth, which is an indication of cache thrashing effects.
More L1 misses draw more data from L2 , but there is headroom in the L2 bandwidth (400 GB/s at 45 processes versus 100 GB/s at 44 processes; see Figure~\ref{fig:singleNodeMEMCaches}(d)), indicating that L2 is not the bottleneck but rather the L1 cache.
Since TLBs are extremely sensitive to alignment problems, several parallel data streams in a Structure of Array (SoA) memory-layout for \CODE{bm} may cause problems, since many concurrent data streams hit different pages, leading to a shortage of TLB entries; L1 cache bank conflicts are also a possible culprit.
Such issues are typically reflected in certain processes being slower if the local domain size is unfortunate. This shows, e.g., in the ITAC timeline for 71 processes on ClusterA (inset of Figure~\ref{fig:singleNodeMEMCaches}(h)), where performance is about 33\% smaller than on 72 processes due to process 70 being significantly slower, leading to extra waiting times on the others.

\highlight{\emph{\textbf{Upshot}}: The SPEChpc 2021 suite covers a wide range of scalability, vectorization, and memory- and non-memo\-ry-bounded\-ness patterns on the node level.
The speedup of Sapphire Rapids vs.\ Ice Lake is within the range expected from peak performance and bandwidth improvements together with larger LLCs.
\CODE{minisweep} suffers an up to 75\% performance hit due to MPI serialization, while \CODE{lbm} shows fluctuating performance with varying process count due to multiple data alignment issues.}

\subsection{CPU and DRAM Power}
\label{sec:NLPower}
In Figure~\ref{fig:singleNodePower} we show CPU and DRAM power for all benchmarks on the node level. In (a) and (c) we choose a representation that allows to identify different scaling patterns and their impact on power dissipation: CPU and DRAM power are plotted against speedup (single-core baseline) up to the first ccNUMA domain boundary (half a socket on ClusterA, a quarter socket on ClusterB). Saturating, scalable, and erratic scaling behavior can be clearly discerned in this way. Even with only a single ccNUMA domain populated, 90\%--95\% of the total power is consumed by the CPU compared to only about 10\% (ClusterA) and 5\% (ClusterB) by the memory modules.
As anticipated and shown in (b) and (d), going from one socket (up to 36 and 52 processes on ClusterA and ClusterB) to two sockets results in a two-fold increase in maximum power. 
In accordance with a naive CPU and DRAM power model, on-chip and DRAM power grows linearly with the number of active cores until a bottleneck is hit, after which additional inactive cores wait for memory, and therefore the slope of on-chip power still grows but more slowly, without additional performance. DRAM power becomes constant after the memory bandwidth has saturated.
However, it also depends on actual memory access pattern, such as continuous vs. burst mode, manufacturing process, DIMM organization, and number of concurrent data streams.

\subsubsection{Hot and cool benchmarks}
There are clearly ``\emph{hot}'' and ``\emph{cool}'' SPEChpc benchmarks with high and low per-CPU power dissipation. The hot benchmarks come close to the TDP of both systems. For instance, on ClusterA and ClusterB, \CODE{sph-exa} achieves 98\% and 97\% of the socket TDP (244~W and 333~W), while \CODE{soma} reaches only 89\% and 85\%  (222~W and 298~W); see Figure~\ref{fig:singleNodePower}.

The low-intensity memory-bound benchmarks \CODE{\{pot3d, tealeaf, cloverleaf\}} attain the highest DRAM power (16~W on one ccNUMA domain on ClusterA and 10-13~W for distinct ccNUMA domain on ClusterB), which remains constant for saturated memory bandwidth and demonstrates how the DRAM's power consumption is strongly tied to the memory bandwidth utilization. 
Conversely, for the high-intensity non-memory-bound \CODE{\{sph-exa, lbm, minisweep, soma\}} benchmarks, most power is drawn in the computational units and the cache hierarchy, 
but their DRAM power is limited, hitting a low of 9.5~W and 5.5~W minimum DRAM power for \CODE{soma} on one ccNUMA domain of ClusterA and ClusterB.

\subsubsection{Fluctuating performance impact on power}
The different power behavior of the \CODE{lbm} and \CODE{minisweep} codes is worth noting. With fluctuating performance, \CODE{lbm} has lower power than \CODE{minisweep}.
Also the performance drops with \CODE{lbm} are associated with slight power drops while in \CODE{minisweep} more cores also lead to more power. This is due to the different reasons for the drops in both codes (MPI waiting time in \CODE{minisweep} vs.\ slow execution in \CODE{lbm}).

\subsubsection{Comparison of system power}
On both systems, the general behavior with respect to power dissipation is similar; see Fig.~\ref{fig:singleNodePower}.
The memory-bound and non-memory-bound benchmarks on ClusterB 
consume 40\% and 25\% more on-chip power than on ClusterA, respectively.
On the two recent architectures considered here, the extrapolated zero-core baseline for chip power is now substantially higher than that of earlier architectures, being around 50\% of the 350~W TDP on Sapphire Rapids (176--181 W) and 40\% of the 250~W TDP on Ice Lake (95--101 W). To put this into perspective, on the Sandy Bridge server architecture from 2012, baseline power only accounted for less than 20\% of the 120~W TDP~\cite{Hager2012,AfzalThesis:2015}.

The DDR5 memory on ClusterB is more power efficient and has significantly less impact on total power than 
DDR4 on ClusterA despite its larger size. It employs a lower voltage and half-rate analog and digital clocking, i.e., DDR5 can achieve the same data transfer rate with half clock frequency~\cite{Nassif2022}.

\highlight{\emph{\textbf{Upshot}}: A part of the SPEChpc 2021 benchmarks attain power dissipation close to TDP.
The latest generation of processors exhibit a substantial idle power and a rather cool DDR5 RAM with half-rate clocking.
}

\subsection{CPU and DRAM Energy to solution}\label{sec:NLE}
Energy to solution, along with the energy-delay product (EDP) is a primary metric for energy efficiency. 
In order to study the relevant parameter space and identify optimal operating points, the \emph{Z-plot}~\cite{AfzalThesis:2015} is most useful (see Fig.~\ref{fig:singleNodeEnergy}(a, b)). It relates energy to the speedup (or performance) of a code, with the amount of resources (number of cores here) as a parameter within a data set. In a Z-plot, horizontal lines mark constant energy, vertical lines mark constant speedup (or performance), and lines through the origin mark constant EDP (the slope being proportional to the EDP).

\subsubsection{Energy and EDP minimums}
\begin{figure}[t]
    \begin{minipage}{0.48\textwidth}
        {\includegraphics[scale=0.312]{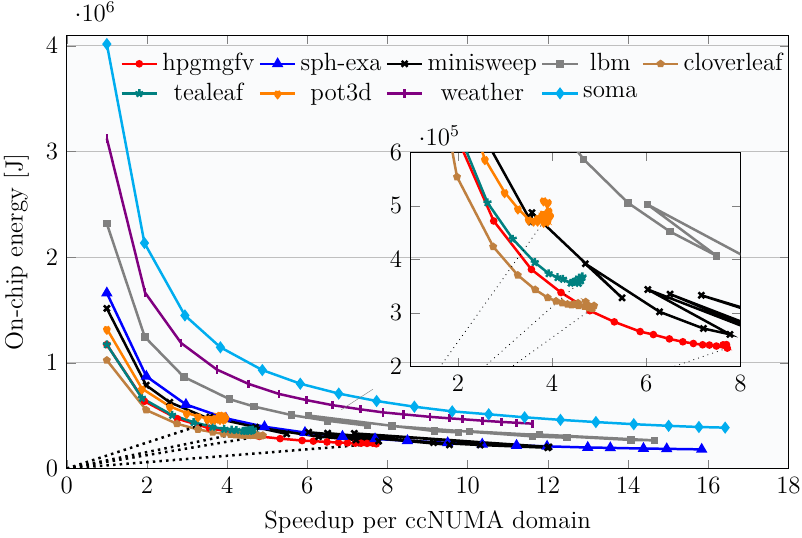}}\hspace{-0.15em}
        {\includegraphics[scale=0.312]{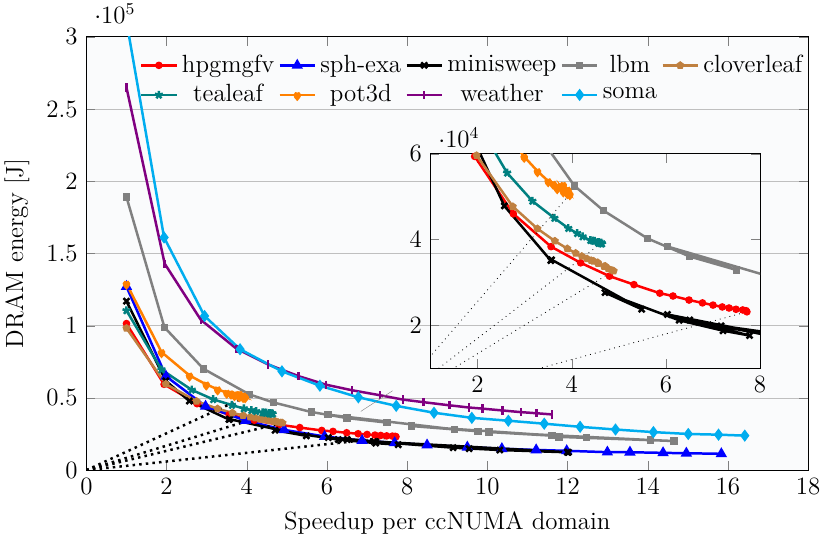}}
        \caption*{(a) ClusterA CPU and DRAM energy (ccNUMA domain)}
        \label{fig:DLEClusterA}
    \end{minipage}%

    \begin{minipage}{0.48\textwidth}
        {\includegraphics[scale=0.312]{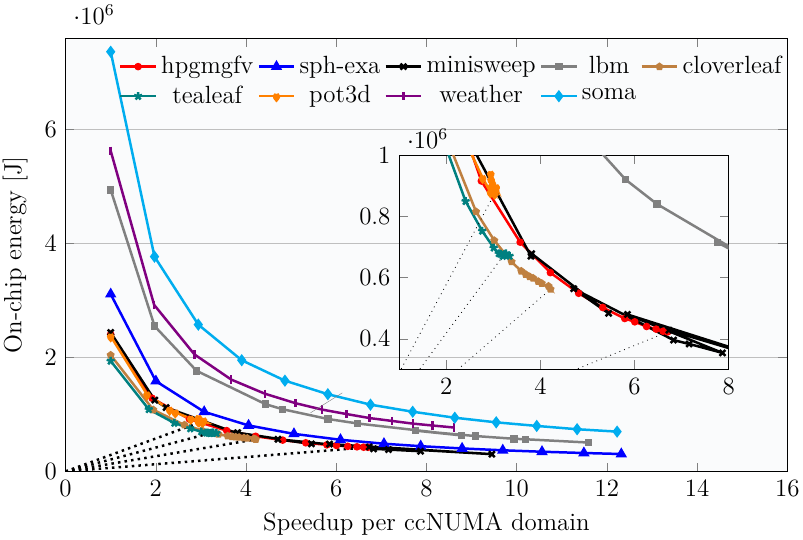}}\hspace{-0.15em}
        {\includegraphics[scale=0.312]{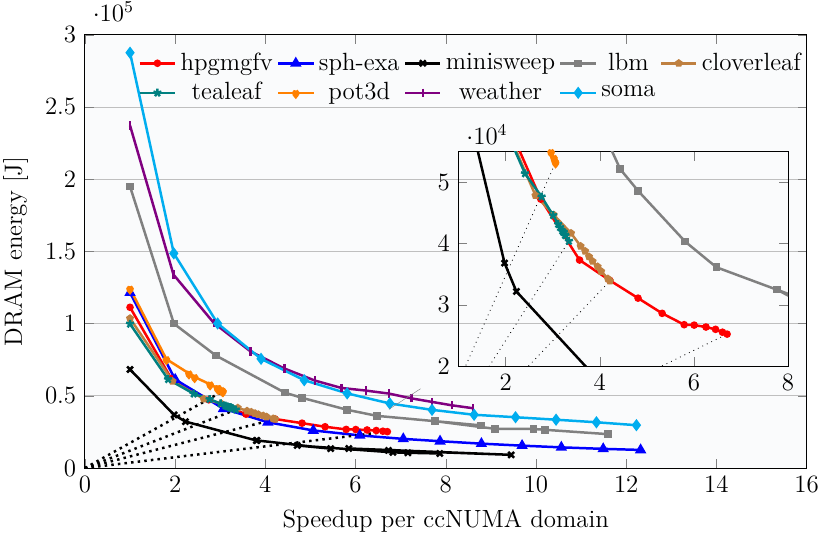}}
        \caption*{(b) ClusterB CPU and DRAM energy (ccNUMA domain)}
        \label{fig:DLEClusterB}
    \end{minipage}%
    
    \begin{minipage}{0.48\textwidth}
        {\includegraphics[scale=0.305]{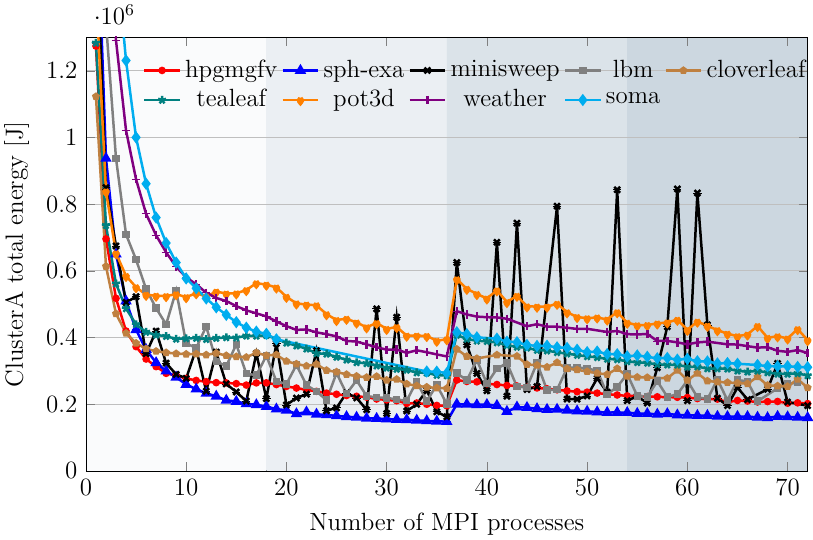}}
        {\includegraphics[scale=0.305]
        {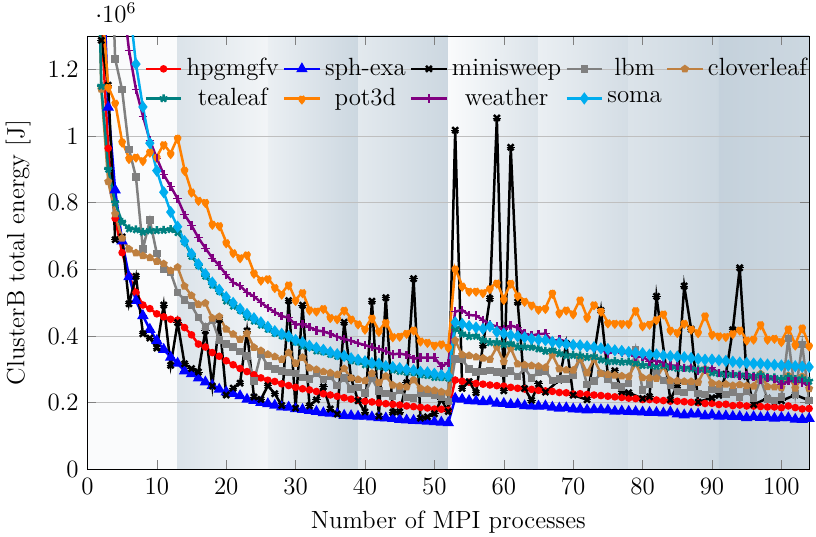}}
        \caption*{(c) ClusterA and ClusterB total energy (node)}
        \label{fig:NLE}
    \end{minipage}%
    \caption{
    SPEChpc 2021 tiny suite energy on both systems.
    (a, b) Z-plot of CPU and DRAM energy vs.\ speedup.
    (c) Total energy (chip and DRAM) vs.\ number of processes.
    }
    \label{fig:singleNodeEnergy}
\end{figure}

In previous Intel architectures, reducing the energy to solution of memory-bound code involved concurrency throttling, i.e., reducing the number of active cores on a ccNUMA domain~\cite{Hager2012,Wittmann2016}.
On the latest designs, however, the baseline power is so dominating that using less than the full ccNUMA domain results in minor energy savings only. Moreover, the minimum energy and minimum EDP operating points are so close together as to be hardly discernible. 
This suggests that making code faster (``code race-to-idle'') is now the primary means of energy reduction. If strong performance fluctuations are present, the race-to-idle rule calls for avoiding low-performance operating points. This is clearly visible in Fig.~\ref{fig:singleNodeEnergy}(c) for the \CODE{lbm} and \CODE{minisweep} codes.

\subsubsection{Comparison of energy across CPUs}
DRAM energy is only a minor contributor to energy to solution. 
If memory bandwidth is the bottleneck, both systems should exhibit comparable energy to solution because the 50\% higher memory bandwidth on ClusterB makes up for the 40\% higher power. 
However, if the core performance is the bottleneck, then the newer CPU would be less efficient because the 40\% increase in power is not offset by a 20\% gain in core performance of Sapphire Rapids over Ice Lake. It goes without saying that the rest of the system (board, network, disks) is ignored in this assessment.
\highlight{\emph{\textbf{Upshot}}: High baseline power on the new Intel CPUs prevents energy from significantly increasing for saturated bandwidth, bringing the $E$ and EDP minimums closer and making runtime equivalent to energy.
On a socket basis, Sapphire Rapids only pays off energy-wise for memory-bound workloads.
}

\section{Multi-node analysis}
\label{sec:MNL}
For large process counts to benefit more from increased workload, the ``small'' workload suite was used for multi-node analysis.

\mypara{Communication routines}
The runtime breakdown obtained using ITAC was consistent with~\cite{Brunst2022}. Under strong scaling, all benchmarks suffer from significant communication overhead.
In decreasing order, \CODE{\{soma, tealeaf, pot3d, sph-exa, cloverleaf, hpgmgfv\}} heavily employ reductions via \CODE{MPI\_Allreduce}. 
Point-to-point communication is the dominant contribution to communication overhead in \CODE{\{weather, minisweep, hpgmgfv, cloverleaf, sph-exa and pot3d\}} (again in decreasing order).
\CODE{MPI\_Barrier} takes significant time in \CODE{lbm}. However, it could be avoided because it is only used to synchronize processes at the end of each iteration.

\subsection{Strong-scaling performance}
\label{sec:MNLP}
The results from multi-node strong scaling experiments shown in Fig.~\ref{fig:multiNodeMEMSpeedup} indicate that two antagonistic effects determine the scaling behavior: communication overhead and memory data volume (specifically that a major part of the working set fits into cache). The following table summarizes which scaling behavior can be attributed to which cause(s) for the different benchmarks:
\begin{table}[h!]
    \centering
	\begin{adjustbox}{width=0.48\textwidth}
            \setlength\extrarowheight{-0.7pt}
            \setlength\tabcolsep{2pt}
            \begin{threeparttable}
            \begin{tabular}[fragile]{zlvzxy}
            	\toprule
            	\rowcolor[gray]{0.9}
                Case & Scalability & Cache effect  & Communication overhead & ClusterA' Codes & ClusterB' Codes\\
            	\midrule
                \cellcolor[gray]{0.9}A & super-linear & $\usebox\HighSpeed$   & $\usebox\LowSpeed$ & \CODE{pot3d} &  \CODE{weather, pot3d}\\
                \cellcolor[gray]{0.9}B & linear &   $\usebox\MediumSpeed$  & $\usebox\MediumSpeed$  & \CODE{weather, tealeaf} & \CODE{tealeaf}\\
                \cellcolor[gray]{0.9}C & close-to-linear & $\usebox\LowSpeed$   & $\usebox\HighSpeed$ & \CODE{hpgmgfv} & \CODE{hpgmgfv} \\
                \cellcolor[gray]{0.9}D & close-to-linear & $\usebox\ZeroSpeed$   & $\usebox\HighSpeed$ & \CODE{cloverleaf} & \CODE{cloverleaf}\\
                \cellcolor[gray]{0.9} & poor & $\usebox\ZeroSpeed$   & $\usebox\HighSpeed$ + small data-set & \CODE{soma\mbox{$^\mathsection$},lbm}  & \CODE{soma\mbox{$^\mathsection$},lbm}\\
                \cellcolor[gray]{0.9}&  &  & & \CODE{ sph-exa, minisweep}  & \CODE{ sph-exa, minisweep}\\
            	\bottomrule
            \end{tabular}
        \begin{tablenotes}
        \small
        \item \mbox{$^\mathsection$} Soma's scaling is constrained by the communication overhead before the excess memory issue becomes problematic.
        
        \noindent\rule[0.5ex]{\linewidth}{1pt}
    \end{tablenotes}
    \end{threeparttable}
	\end{adjustbox}
\end{table}

In Fig.~\ref{fig:multiNodeMEMSpeedup}(b, e) we show the per-node memory bandwidth. Perfect performance scaling with no cache effects would show as a horizontal line here. All benchmarks except \CODE{soma} exhibit a declining per-node bandwidth, which either indicates poor scaling due to communication overhead (or load imbalance) or a combination of cache effects which reduce the memory data volume. This is why we also show the overall memory transfer volume in Fig.~\ref{fig:multiNodeMEMSpeedup}(c, f). 
A rise in data traffic over the optimal horizontal line signals effects such as data replication. 
The similar tendency in code scaling for node-level ``tiny'' and cluster-level ``small'' workloads is caused by a mild change in the problem size per node as both are increased.

\subsubsection{Four fundamental scaling patterns}
In the following we describe the different patterns observed in the multi-node scaling.

\mypara{\textbf{Case A}: Cache effect prevails over communication overhead}
With increasing node count, \CODE{weather} shows a significant decrease in memory data volume and bandwidth on both clusters (but stronger on ClusterB). This cache effect dominates and leads to superlinear scaling. 
The difference between the clusters is due to the fact that ClusterB has 1.45 times more L3 per core and 1.6 times more L2 cache per core than ClusterA, so the working set of \CODE{weather} can fit earlier into the cache of ClusterB's CPUs. 

\mypara{\textbf{Case B}: Communication overhead and cache effects balance out}
The superlinear scaling due to cache effects (Case A) can be counteracted by increasing communication volume or synchronization overhead, to the point of causing linear scaling. The codes \CODE{weather} on ClusterA and  \CODE{tealeaf} on both systems fall under this category.

\mypara{\textbf{Case C}: Communication overhead dominates over cache effect}
In this case, memory traffic drops with increasing node count, but the anticipated super-linear scaling is outweighed by high communication overhead.
On the cluster level, \CODE{hpgmgfv} falls under category. The increasing communication cost is caused by point-to-point communication and reductions.

\begin{figure}[t]
    \begin{minipage}{0.48\textwidth}
        \subfloat[ClusterA speedup]
        {\includegraphics[scale=0.311]{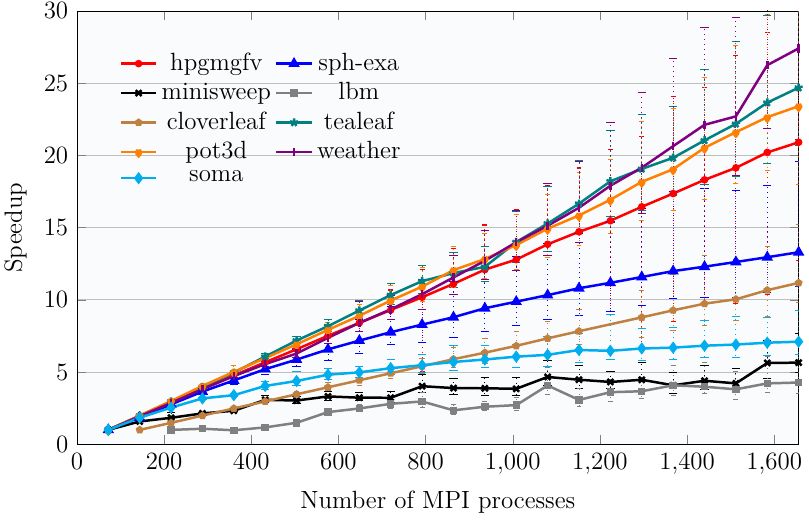}}\hspace{0em}
        \subfloat[ClusterB speedup]
        {\includegraphics[scale=0.311]{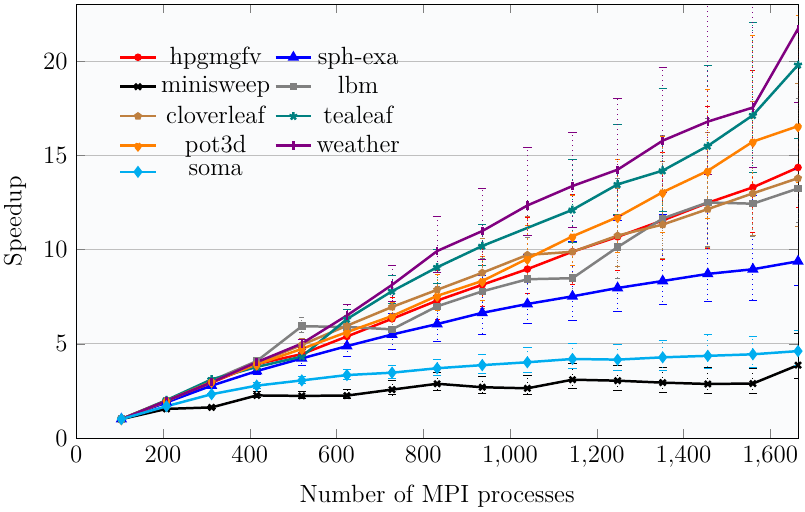}}
    \end{minipage}%

    \begin{minipage}{0.48\textwidth}
        \subfloat[ClusterA memory bandwidth]
        {\includegraphics[scale=0.305]{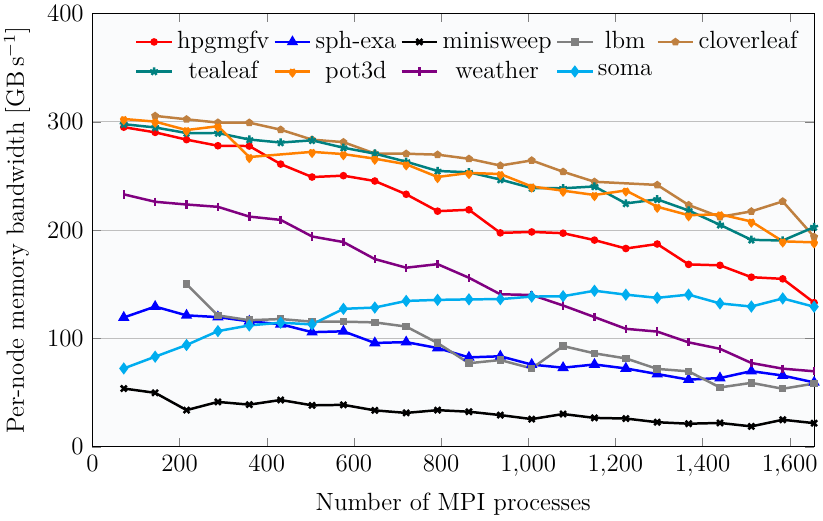}}\hspace{0.05em}
        \subfloat[ClusterB memory bandwidth]
        {\includegraphics[scale=0.305]{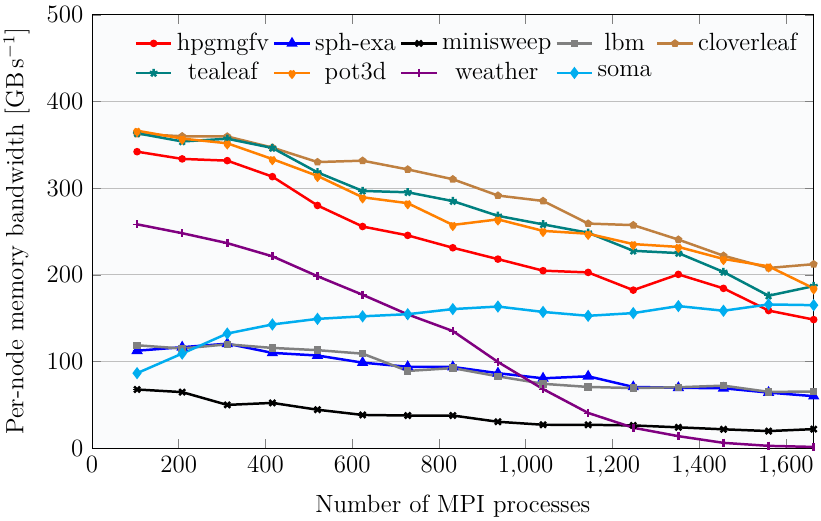}}
    \end{minipage}%
    
    \begin{minipage}{0.48\textwidth}
        \subfloat[ClusterA memory data volume]
        {\includegraphics[scale=0.305]{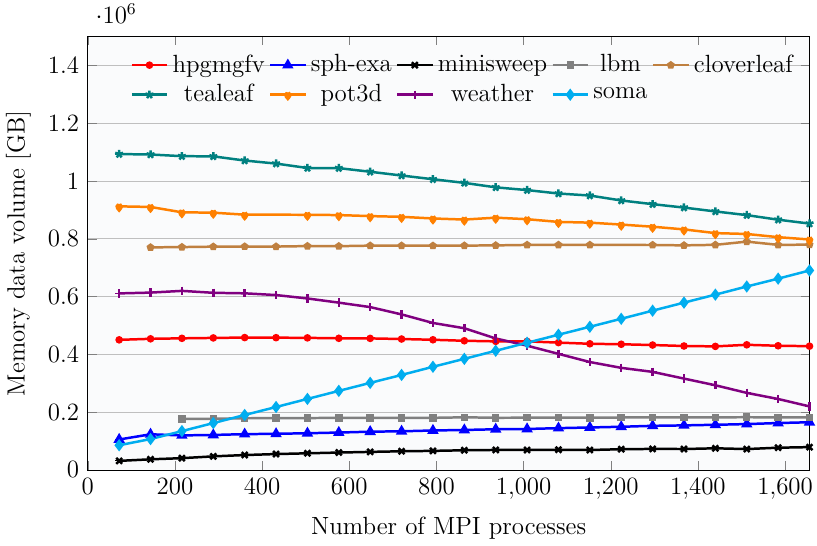}}\hspace{0.05em}
         \subfloat[ClusterB memory data volume]
        {\includegraphics[scale=0.305]{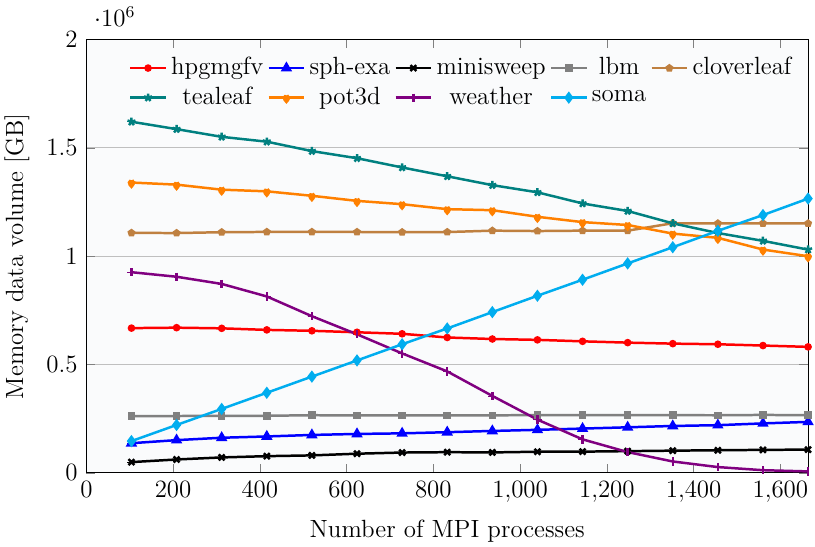}}
    \end{minipage}%
    \caption{
    SPEChpc 2021 small suite scaling on ClusterA (left) and ClusterB (right).
    (a, d) Speedup with min, max and average statistics,
    (b, e) per-node memory bandwidth, and
    (c, f) aggregate data volume. 
    }
    \label{fig:multiNodeMEMSpeedup}
\end{figure}

\mypara{\textbf{Case D}: No cache effect; only communication overhead}
In this case, significant MPI communication is the only factor contributing to the poor scaling, where the memory data volume remains the same and the memory bandwidth declines. The \CODE{cloverleaf} and \CODE{soma} codes on both systems fall under this category.
Further, data set size affects communication overhead, i.e., smaller data sets have a higher likelihood of having significant communication overhead.
The poor scaling observed in \CODE{\{minisweep, soma, sph-exa\}} is caused by a confluence of large MPI communication \{blocking pair-wise, \CODE{MPI\_Allreduce}, both\} and a comparatively small data set; see Fig.~\ref{fig:multiNodeMEMSpeedup}(c, f). 

\subsubsection{Intriguing non-memory-bound case of soma}
Out of the nine benchmarks, \CODE{soma} is the one that spends the majority of its total time in MPI reductions.
Beyond one to three nodes, \CODE{soma} does not scale effectively. 
However, with increasing node count it draws significant and increasing memory bandwidth, up to about half the maximum on ClusterA and up to 33\% of the maximum on ClusterB.
Hence, the memory bandwidth per node increases while scaling is poor, which constitutes an unusual pattern.
The memory data volume sheds more light on this issue:
A perfect linear rise in aggregated data traffic vs.\ number of nodes on both systems indicates that \CODE{soma} appears to have a lot of replicated data; as long as the code remains non-memory-bound, this might be of minor significance. However, given a linear rise in memory traffic and a logarithmic rise in reduction overhead, the question arises whether at some point (i.e., number of nodes) the code might become memory bound. 
Our measurements demonstrate that this can not happen with \CODE{soma} at least within the ``small'' working set and number of processes for which it is officially designed:
The per-node memory bandwidth initially increases to 150~\GBS\ (far below the limit of 350~\GBS), and then remains essentially constant at 150~\GBS, at which point the code ceases scale at all. 

\subsubsection{Comparison of cluster performance}
The two clusters' interconnects are identical so differences in communication performance are not expected. 
The fundamental distinction of ClusterB's node is that it features more cores, more cache per core, higher memory bandwidth, and higher machine balance (memory bandwidth to peak ratio).  
Our findings reveal that the tendency in code scaling, whether it is poor or good, is qualitatively consistent across clusters.
However, the superlinear multi-node scaling of \CODE{weather} is stronger on ClusterB due to its larger cache.  The scaling of \CODE{cloverleaf} is slightly worse on ClusterB due to the higher single-node baseline (250~\GFS\ vs.\ 160~\GFS, owing to memory bandwidth).
Similarly, \CODE{sph-exa}'s significantly inferior scaling efficiency on ClusterB is caused by the 47\% higher performance on a single node compared to ClusterA (6.2~\GFS\ vs.\ 4.2~\GFS).

\highlight{\emph{\textbf{Upshot}}: Based on two antagonistic effects, namely communication overhead and cache effects (measured using fundamental resource metrics), all SPEChpc 2021 benchmarks fall under four fundamentally distinct categories. Especially interesting cases are \CODE{soma} with its excess memory traffic with rising process count and \CODE{weather} with its strongly superlinear scaling.}

\subsection{Scaling impact on power and energy}
\label{sec:MNLE}
\begin{figure}[t!]
    \centering
    \begin{minipage}{0.24\textwidth}
    \hspace{-0.8em}
        \centering
        {\includegraphics[scale=0.32]{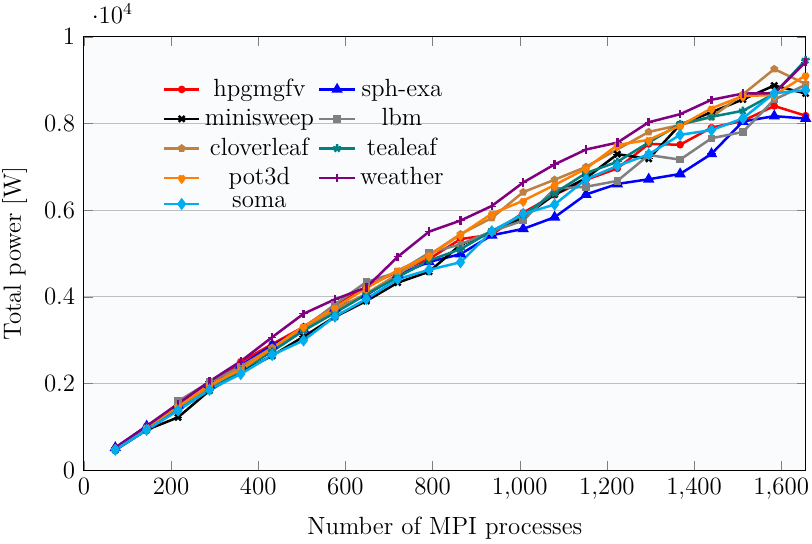}}
        \caption*{(a) ClusterA total power}
        \label{fig:MNPClusterA}
    \end{minipage}%
    \begin{minipage}{0.24\textwidth}
        \centering
        {\includegraphics[scale=0.32]{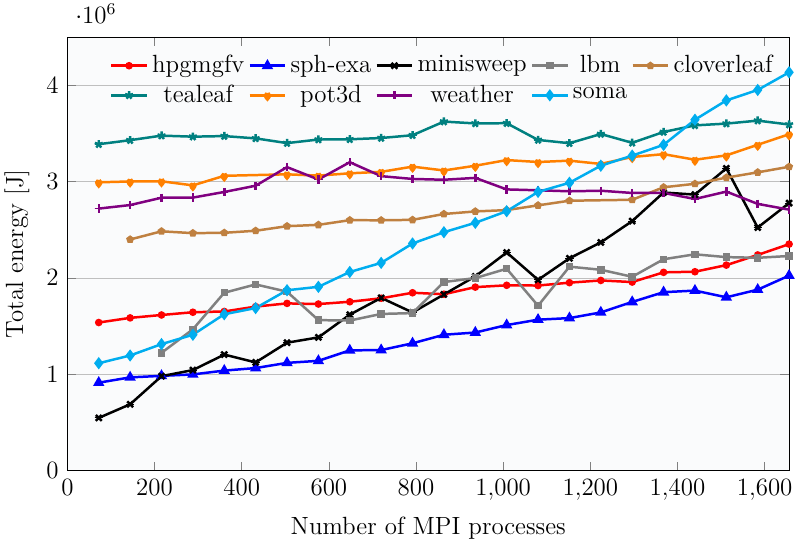}}
        \caption*{(b) ClusterA total energy}
        \label{fig:MNEClusterA}
    \end{minipage}%
    
    \begin{minipage}{0.24\textwidth}
    \hspace{-0.5em}
        \centering
        {\includegraphics[scale=0.315]
        {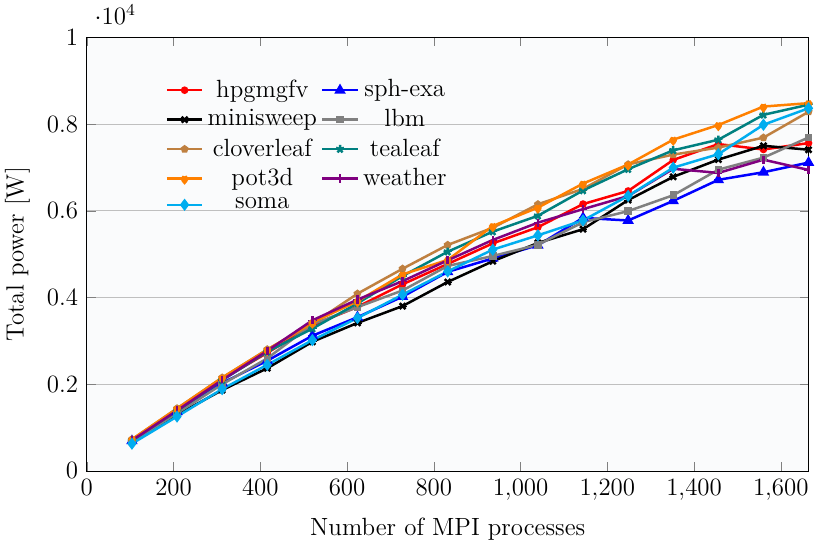}}
        \caption*{(c) ClusterB total power}
        \label{fig:MNPClusterB}
    \end{minipage}%
    \begin{minipage}{0.24\textwidth}
        \centering
        {\includegraphics[scale=0.315]
        {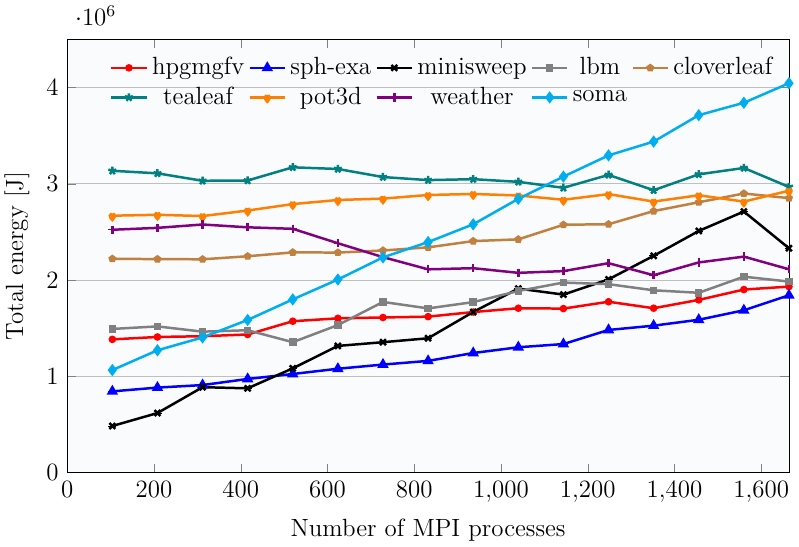}}
        \caption*{(d) ClusterB total energy}
        \label{fig:MNEClusterB}
    \end{minipage}%
    \caption{SPEChpc 2021 small suite 
    total (chip+DRAM) power and energy scaling on multiple nodes for ClusterA (top) and ClusterB (bottom).
    }
    \label{fig:multiNodePowerEnergy}
\end{figure}

Figure~\ref{fig:multiNodePowerEnergy}(a, c) shows total power dissipation scaling on multiple nodes. The codes of the suite attain 74--85\% (ClusterA) and 63--76\% (ClusterB) of the CPU TDP limit on the full set of nodes (5.9--6.8~kW out of 8~kW on ClusterA and 7.1--8.5~kW out of 11.2~kW on ClusterB).
The baseline power of the coolest code dominates the dynamic power, with a share of 82\% (5.8~KW vs.\ 1.3~KW) on ClusterB and 53\%\ on ClusterA (3.1~KW vs.\ 2.8~KW). 
The poorly scalable benchmarks \CODE{\{minisweep, soma, sph-exa\}} require more resources (node-hours) with increasing node count, which leads to rising energy consumption (see Fig.~\ref{fig:multiNodePowerEnergy}(b, d)).
Scalable codes such as \CODE{tealeaf} exhibit constant energy as anticipated.
For \CODE{soma}, the overall energy rises linearly up to three nodes but with a steeper slope beyond because of declining scalability. 

\highlight{\emph{\textbf{Upshot}}: Multi-node energy consumption scaling  mainly depends on the scaling properties of a code; poorly scalable benchmarks always burn more energy when scaling out, with \CODE{soma} marking an especially interesting case. }

\section{Summary and future work}\label{sec:conclusion}

We provided an in-depth node-level and multi-node analysis of the MPI-only versions of the SPEChpc 2021 benchmarks with respect to power/energy and performance on clusters based on Intel Ice Lake and Sapphire Rapids CPUs. Using speedup, performance, data traffic, and bandwidth metrics we could categorize the codes with respect to their memory and communication boundedness, and we could uncover unusual scaling patterns that are rooted in excess replicated data, performance bugs in MPI communication, data alignment issues, and, in one case, counteracting superlinear scaling and communication overhead effects. One the node level, the overall performance advantage of Sapphire Rapids vs.\ Ice Lake is in line with expectations based on maximum memory bandwidth, peak performance, and cache size ratios.

We showed that there is a 25\% variation in power dissipation on the package level across benchmarks, but that the ``hot'' codes are able to come very close to the TDP limit on both CPUs. We also observed that idle power, i.e., the hypothetical power dissipation of the CPU with zero active cores, takes a much higher fraction of the overall power than in older CPUs. This has the important consequence that minimum energy to solution and energy-delay product operating points are practically identical, idling cores save negligible energy, and race-to-idle via code optimization becomes the pivotal energy reduction strategy. On Sapphire Rapids, DRAM power is measurably lower than on Ice Lake due to the new DDR5 technology; however, this reduction is all but insignificant considering the chip and full-system power dissipation. 

In future work we will more thoroughly investigate optimization opportunities and further performance patterns (such as the multi-faceted fluctuations in \CODE{lbm}) as well as further parallelization approaches beyond pure MPI. Furthermore, we expect insight from studying  desynchronization~\cite{AfzalHW:2020,AfzalHW:2022:1,AfzalHW:2022:2,AfzalHW:2022:4,AfzalHW:2023:1} and idle wave phenomena~\cite{AfzalHW:2019,AfzalHW:2021} in those benchmarks that show  mixture of memory-, compute-, and communication-bound behavior. 
 
\ifblind
\else
\section*{Acknowledgments}
The authors gratefully acknowledge the scientific support and HPC resources provided by the Erlangen National High Performance Computing Center (NHR@FAU) of the Friedrich-Alexander-Universität Erlangen-Nürnberg (FAU) and LRZ Gar\-ching. NHR funding is provided by federal and Bavarian state authorities. The hardware at NHR@FAU is partially funded by the German Research Foundation (DFG) -- 440719683. This work was partly supported by the Competence Network for Scientific High-Performance Computing in Bavaria (KONWIHR) under project ``OMI4Papps.'' 
\fi
\balance
\bibliographystyle{ACM-Reference-Format}
\bibliography{references_woDOI}

\end{document}

%% file: setup.tex
\AtBeginDocument{%
  \providecommand\BibTeX{{%
    \normalfont B\kern-0.5em{\scshape i\kern-0.25em b}\kern-0.8em\TeX}}}

\copyrightyear{2023}
\acmYear{2023}
\acmDOI{XXXXXXX.XXXXXXX}
\acmConference[SC '23]{The International Conference for High Performance Computing, Networking, Storage, and Analysis}{November 14--19, 2023}{Denver, CO, USA}
\acmBooktitle{SC '23: The International Conference for High Performance Computing, Networking, Storage, and Analysis, November 14--19, 2023, Denver, CO, USA} 
\acmPrice{XX.00}
\acmISBN{978-1-4503-XXXX-X/18/06}

\setcopyright{none}
\usepackage{acro}
\usepackage{adjustbox}
\usepackage{xtab,booktabs}
\usepackage{comment}
\usepackage{todonotes}
\usepackage{colortbl}
\usepackage[justification=centering]{subcaption} 
\usepackage{listings}
\usepackage[para,online,flushleft]{threeparttable}
\usepackage{longtable}

\newcolumntype{B}{>{\columncolor{cyan!4!white}}p{0.46\textwidth}}
\newcolumntype{a}{>{\columncolor{gray!10!white}}c}
\newcolumntype{x}{>{\columncolor{green!10!white}}c}
\newcolumntype{y}{>{\columncolor{blue!10!white}}c}
\newcolumntype{z}{>{\columncolor{yellow!10!white}}c}
\newcolumntype{v}{>{\columncolor{red!10!white}}c}
\definecolor{OliveGreen}{rgb}{0,0.6,0}
\definecolor{ForestGreen}{RGB}{34,139,34}
\definecolor{myblue}{RGB}{37,165,203}
\definecolor{FAUblue}{rgb}{0.000, 0.2196, 0.3961}
\definecolor{myred}{RGB}{175,32,67}

\newcommand{\bq}{\begin{equation}}
\newcommand{\eq}{\end{equation}}

\newcommand{\second}{\mbox{s}}

\newcommand{\flop}{\mbox{flop}}

\newcommand{\bit}{\mbox{bit}}

\newcommand{\GBPS}{\mbox{G\bit/\second}}

\newcommand{\GBS}{\mbox{GB/\second}}

\newcommand{\GFS}{\mbox{G\flop/\second}}

\newcommand{\GHZ}{\mbox{GHz}}

\newcommand{\GB}{\mbox{GB}}
\newcommand{\TB}{\mbox{TB}}

\newcommand{\GiB}{\mbox{GiB}}
\newcommand{\MiB}{\mbox{MiB}}
\newcommand{\KiB}{\mbox{KiB}}

\newcommand{\LLC}{last-level cache\xspace}

\lstdefinestyle{small}{
  language=C++,
  basicstyle=\footnotesize\tt,
  showspaces=false,
  showstringspaces=false,
  breakindent=0pt,
  breaklines,
  captionpos=b,
  breakatwhitespace=true,
  numbers=left,
  numberstyle=\tiny,
  stepnumber=2,
  numbersep=7pt,
  linewidth=0.47\textwidth,
  xleftmargin=.02\textwidth, 
  xrightmargin=.02\textwidth,
  keywordstyle=\color{blue},
  tabsize=2,
  backgroundcolor=\color{white!97!black},
}
\lstdefinestyle{smallWithColor}{
  basicstyle=\small\tt,
  language=C++,
  showspaces=false,
  showstringspaces=false,
  breakindent=0pt,
  breaklines,
  captionpos=b,
  breakatwhitespace=true,
  numbers=left,
  numberstyle=\tiny,
  stepnumber=1,
  numbersep=7pt,
  keywordstyle=\color{blue},
  tabsize=2,
  moredelim=**[is][\color{red}]{@}{@},
  moredelim=**[is][\color{green}]{|}{|},
  backgroundcolor=\color{white!97!black},
}

\lstdefinestyle{footnotesizeStyle}{
  basicstyle=\footnotesize\tt,
  language=C++,
  showspaces=false,
  showstringspaces=false,
  breakindent=0pt,
  breaklines,
  captionpos=b,
  breakatwhitespace=true,
  numbersep=5pt,
  keywordstyle=\color{blue},
  tabsize=2,
  moredelim=**[is][\color{red}]{@}{@},
  moredelim=**[is][\color{green}]{|}{|},
  backgroundcolor=\color{white!97!black},
}

\usepackage[htt]{hyphenat} 
  
\newcommand{\enquote}[1]{``#1''}  
\newcommand{\CODE}[1]{\texttt{#1}}
\newcommand{\mypara}[1]{\emph{#1}.~}

\colorlet{backgroundcol}{cyan!10!white}
\newcommand{\highlight}[1]{%
	\par\noindent
	\fcolorbox{black}{backgroundcol}{%
		\parbox{\dimexpr\linewidth-2\fboxsep\relax}{%
			#1
		}%
}}

\colorlet{backgroundcolumn}{yellow!10!white}

\newcommand{\acrodef}[2]{\DeclareAcronym{#1}{short={#1},long={#2}}}
\acrodef{CER}{commu\-ni\-cation-to-exe\-cu\-tion ratio}
\acrodef{ChebFD}{Chebyshev filter diagonalization}
\acrodef{EOS}{Equation of State}
\acrodef{HPCG}{High Performance Conjugate Gradient}
\acrodef{LBM}{Lattice Boltzmann Menthod}
\acrodef{LULESH}{Livermore Unstructured Lagrangian Explicit Shock Hydrodynamics}
\acrodef{MG}{multigrid}
\acrodef{MST}{MPI-augmented STREAM Triad}
\acrodef{SPEC}{Standard Performance Evaluation Corporation}
\acrodef{SoA}{structure of arrays}
\acrodef{SpMVM}{sparse matrix-vector multiplication}

\usepackage{pgfplots}
\usepackage{cancel}
\usepackage{pgfplotstable}
\usepackage{multirow}
\usepgfplotslibrary{colorbrewer}
\usetikzlibrary{pgfplots.statistics, pgfplots.colorbrewer} \usepackage{forest}
\usetikzlibrary{arrows.meta, shapes.geometric, calc, shadows}
\usetikzlibrary{fit,calc}
\usepackage{smartdiagram}
\usepackage[ruled]{algorithm2e}

\colorlet{mygreen}{green!75!black}
\colorlet{col1in}{red!30}
\colorlet{col1out}{red!40}
\colorlet{col2in}{mygreen!40}
\colorlet{col2out}{mygreen!50}
\colorlet{col3in}{blue!30}
\colorlet{col3out}{blue!40}
\colorlet{col4in}{mygreen!20}
\colorlet{col4out}{mygreen!30}
\colorlet{col5in}{blue!10}
\colorlet{col5out}{blue!20}
\colorlet{col6in}{blue!20}
\colorlet{col6out}{blue!30}
\colorlet{col7out}{orange}
\colorlet{col7in}{orange!50}
\colorlet{col8out}{orange!40}
\colorlet{col8in}{orange!20}
\colorlet{linecol}{blue!60}
\usepgfplotslibrary{colorbrewer}

\usetikzlibrary{decorations.pathmorphing}
\tikzset{discont/.style={decoration={zigzag,segment length=1.5pt, amplitude=4pt},decorate}}
\def\discontarrow(#1)(#2)(#3)(#4);{
	\draw[discont] (#2) -- (#3);
	\draw[] (#1) -- (#2) (#3) -- (#4);
}

\usepgflibrary{decorations.fractals}
\pgfplotstableset{
rowfont/.style={
	postproc cell content/.append code={
		\count0=\pgfplotstablerow
		\advance\count0 by1
		\ifnum\count0=#1
		\pgfkeysalso{@cell content/.add={\bfseries\boldmath\color{red}}{}}
		\fi
	},
},
colfont/.style={
	postproc cell content/.append code={
		\count0=\pgfplotstablecol
		\advance\count0 by1
		\ifnum\count0=#1
		\pgfkeysalso{@cell content/.add={\color{black}}{}}
		\fi
	},
},
colfontred/.style={
	postproc cell content/.append code={
		\count0=\pgfplotstablecol
		\advance\count0 by1
		\ifnum\count0=#1
		\pgfkeysalso{@cell content/.add={\bfseries\boldmath\color{red}}{}}
		\fi
	},
},
colfontwhite/.style={
	postproc cell content/.append code={
		\count0=\pgfplotstablecol
		\advance\count0 by1
		\ifnum\count0=#1
		\pgfkeysalso{@cell content/.add={\bfseries\boldmath\color{white}}{}}
		\fi
	},
},	
}

\usepackage{colortbl}
\usepgfplotslibrary{colormaps}
\pgfplotsset{compat=1.17}
\pgfplotsset{colormap/Set3} 
\usepackage{cellspace}
\newcolumntype{C}{}
\newlength{\cellspacelimit}
\pgfplotstableset{
	/color cells/min/.initial=0,
	/color cells/max/.initial=100,
	/color cells/textcolor/.initial=,
	color cells/.code={%
		\pgfqkeys{/color cells}{#1}%
		\pgfkeysalso{%
			postproc cell content/.code={%
				\begingroup
				\pgfkeysgetvalue{/pgfplots/table/@preprocessed
					cell content}\value
				\ifx\value\empty
				\endgroup
				\else
				\pgfmathfloatparsenumber{\value}%
				\pgfmathfloattofixed{\pgfmathresult}%
				\let\value=\pgfmathresult
				\pgfplotscolormapaccess
				[\pgfkeysvalueof{/color cells/min}:\pgfkeysvalueof{/color
					cells/max}]
				{\value}
				{\pgfkeysvalueof{/pgfplots/colormap name}}%
				\pgfkeysgetvalue{/pgfplots/table/@cell content}\typesetvalue
				\pgfkeysgetvalue{/color cells/textcolor}\textcolorvalue
				\toks0=\expandafter{\typesetvalue}%
				\xdef\temp{%
					\noexpand\pgfkeysalso{%
						@cell content={%
							\noexpand\cellcolor[rgb]{\pgfmathresult}%
							\noexpand\definecolor{mapped
								color}{rgb}{\pgfmathresult}%
							\ifx\textcolorvalue\empty
							\else
							\noexpand\color{\textcolorvalue}%
							\fi
							\the\toks0 %
						}%
					}%
				}%
				\endgroup
				\temp
				\fi
			}%
		}%
	},
	every column/.style={column type={Cc!{\color{white}\vrule width 1pt}}},
	every first column/.style={reset styles, column type={l}, string type},
	every head row/.style={before row=\toprule, after row=\midrule},
	after row={\noalign{\global\arrayrulewidth=1pt}\arrayrulecolor{white}\hline},
	every last row/.style={after row=\arrayrulecolor{black}\bottomrule}, 
	@content options for rows/.style 2 args={
		postproc cell content/.add code={%
			\def\isInInputList{0}%
			\pgfplotsforeachungrouped \II in {#1} {%
				\ifnum\II<0
				\count0=\II
				\advance\count0 by\pgfplotstablerows
				\edef\II{\the\count0 }%
				\fi
				\ifnum\II=\pgfplotstablerow\relax
				\def\isInInputList{1}%
				\fi
			}%
			\if1\isInInputList%
			\pgfkeysalso{#2}%
			\fi
		}{},
	},
}

\pgfkeys{/pgf/number format/.cd,1000 sep={\,},fixed,precision=3}

\makeatletter
\pgfplotsset{
	boxplot/hide outliers/.code={
		\def\pgfplotsplothandlerboxplot@outlier{}%
	}
}
\makeatother

\usepgfplotslibrary{units}
\usetikzlibrary{arrows, calc, spy, patterns, backgrounds,decorations.pathreplacing, arrows.meta}

\pgfplotsset{
	my ybar legend/.style={
		legend image code/.code={
			\draw [##1] (0cm,-0.6ex) rectangle +(1.75em,1.1ex);
		},
	},
}

\makeatletter
\pgfdeclarepatternformonly[\hatchdistance,\hatchthickness]{flexible hatch}
{\pgfqpoint{0pt}{0pt}}
{\pgfqpoint{\hatchdistance}{\hatchdistance}}
{\pgfpoint{\hatchdistance-1pt}{\hatchdistance-1pt}}%
{
	\pgfsetcolor{\tikz@pattern@color}
	\pgfsetlinewidth{\hatchthickness}
	\pgfpathmoveto{\pgfqpoint{0pt}{0pt}}
	\pgfpathlineto{\pgfqpoint{\hatchdistance}{\hatchdistance}}
	\pgfusepath{stroke}
}

\makeatletter
\pgfplotsset{
	discontinuous line/.code={
		\pgfkeysalso{mesh, shorten <=#1, shorten >=#1,
			legend image code/.code={
				\draw [##1, shorten <=0cm] (0cm,0cm) -- (0.3cm,0cm);
				\draw [only marks] plot coordinates {(0.3cm,0cm)};
				\draw [##1, shorten >=0cm] (0.3cm,0cm) -- (0.6cm,0cm);
		}}
		\def\pgfplotsplothandlermesh@VISUALIZE@std@fill@andor@stroke{%
			\pgfplotspatchclass{\pgfplotsplothandlermesh@patchclass}{fill path}%
			\pgfplotsplothandlermesh@definecolor
			\pgfusepath{stroke}
			\pgfplotsplothandlermesh@show@normals@if@configured
		}%
	},
	discontinuous line/.default=1.5mm
}
\makeatother

\usetikzlibrary{patterns}
	\newlength{\hatchspread}
	\newlength{\hatchthickness}
	\newlength{\hatchshift}
	\newcommand{\hatchcolor}{}
	\tikzset{hatchspread/.code={\setlength{\hatchspread}{#1}},
		hatchthickness/.code={\setlength{\hatchthickness}{#1}},
		hatchshift/.code={\setlength{\hatchshift}{#1}},
		hatchcolor/.code={\renewcommand{\hatchcolor}{#1}}}
	\tikzset{hatchspread=3pt,
		hatchthickness=0.4pt,
		hatchshift=0pt,
		hatchcolor=black}
	\pgfdeclarepatternformonly[\hatchspread,\hatchthickness,\hatchshift,\hatchcolor]
	{custom north west lines}
	{\pgfqpoint{\dimexpr-2\hatchthickness}{\dimexpr-2\hatchthickness}}
	{\pgfqpoint{\dimexpr\hatchspread+2\hatchthickness}{\dimexpr\hatchspread+2\hatchthickness}}
	{\pgfqpoint{\dimexpr\hatchspread}{\dimexpr\hatchspread}}
	{
		\pgfsetlinewidth{\hatchthickness}
		\pgfpathmoveto{\pgfqpoint{0pt}{\dimexpr\hatchspread+\hatchshift}}
		\pgfpathlineto{\pgfqpoint{\dimexpr\hatchspread+0.15pt+\hatchshift}{-0.15pt}}
		\ifdim \hatchshift > 0pt
		\pgfpathmoveto{\pgfqpoint{0pt}{\hatchshift}}
		\pgfpathlineto{\pgfqpoint{\dimexpr0.15pt+\hatchshift}{-0.15pt}}
		\fi
		\pgfsetstrokecolor{\hatchcolor}
		\pgfusepath{stroke}
	}
	\pgfdeclarepatternformonly[\hatchspread,\hatchthickness,\hatchshift,\hatchcolor]
	{custom north east lines}
	{\pgfqpoint{\dimexpr-2\hatchthickness}{\dimexpr-2\hatchthickness}}
	{\pgfqpoint{\dimexpr\hatchspread+2\hatchthickness}{\dimexpr\hatchspread+2\hatchthickness}}
	{\pgfqpoint{\dimexpr\hatchspread}{\dimexpr\hatchspread}}
	{
		\pgfsetlinewidth{\hatchthickness}
		\pgfpathmoveto{\pgfqpoint{\dimexpr\hatchshift-0.15pt}{-0.15pt}}
		\pgfpathlineto{\pgfqpoint{\dimexpr\hatchspread+0.15pt}{\dimexpr\hatchspread-\hatchshift+0.15pt}}
		\ifdim \hatchshift > 0pt
		\pgfpathmoveto{\pgfqpoint{-0.15pt}{\dimexpr\hatchspread-\hatchshift-0.15pt}}
		\pgfpathlineto{\pgfqpoint{\dimexpr\hatchshift+0.15pt}{\dimexpr\hatchspread+0.15pt}}
		\fi
		\pgfsetstrokecolor{\hatchcolor}
		\pgfusepath{stroke}
	}

\usepackage{awesomebox}
\usepackage{lipsum}
\usepackage{tikz}
\tikzset{pics/speedometer/.style={code={
 \foreach \X/\Y [count=\Z] in {green!70!black/low,orange/medium,red/high}
  {\fill[fill=\X] (240-\Z*60:4) arc(240-\Z*60:180-\Z*60:4) -- 
    (180-\Z*60:3) arc(180-\Z*60:240-\Z*60:3) -- cycle;}
   \fill (180-#1+8:0.3) arc (180-#1+8:180-#1+344:0.3) -- (180-#1-0.5:3.25)
   -- (180-#1+0.5:3.25) --  cycle;
  }}}
\newsavebox\ZeroSpeed  
\newsavebox\LowSpeed  
\newsavebox\MediumSpeed  
\newsavebox\HighSpeed  
\sbox\ZeroSpeed{\scalebox{0.07}{\tikz{\pic{speedometer=0};}}}
\sbox\LowSpeed{\scalebox{0.07}{\tikz{\pic{speedometer=45};}}}
\sbox\MediumSpeed{\scalebox{0.07}{\tikz{\pic{speedometer=90};}}}
\sbox\HighSpeed{\scalebox{0.07}{\tikz{\pic{speedometer=135};}}}

%% file: figures/SPEChpc2021.tex
\begin{table}
	\centering
    \caption{Key attributes of SPEChpc 2021 parallel benchmarks. 
    }
    \label{tab:SPEChpc2021}
    \centering
    \begin{adjustbox}{width=0.48\textwidth}
    \begin{threeparttable}
    \begin{tabular}[fragile]{ccccc>{~}lvx}
		\toprule
        \rowcolor[gray]{0.9}
        \rowcolor[gray]{0.9}
		&&&&&& Tiny   &  Small \\
         \rowcolor[gray]{0.9}
		&&&&&& 5ID.Name\_t   &  6ID.Name\_s \\
  		\rowcolor[gray]{0.9}
        &&&&&Input configuration & 0--64~\GB\ memory   &  0--480~\GB\ memory  \\
		\rowcolor[gray]{0.9}
        \multirow{-4}{*}{\rotatebox{90}{\cellcolor[gray]{0.9} Name}}&\multirow{-4}{*}{\rotatebox{90}{\cellcolor[gray]{0.9} ID}}&\multirow{-4}{*}{\rotatebox{90}{\cellcolor[gray]{0.9} Language}}&\multirow{-4}{*}{\rotatebox{90}{\cellcolor[gray]{0.9} LOC}}&\multirow{-4}{*}{\rotatebox{90}{\cellcolor[gray]{0.9} Collective}}&& 1--256 processes  &  64--1024 processes  \\
		\midrule
		\cellcolor[gray]{0.9}&\cellcolor[gray]{0.9}&\cellcolor[gray]{0.9}&\cellcolor[gray]{0.9}&\cellcolor[gray]{0.9}&\{X,Y\}-dimension of lattice  & \{4096,16384\}   & \{12000,48000\}    \\    
		\cellcolor[gray]{0.9}&\cellcolor[gray]{0.9}&\cellcolor[gray]{0.9}&\cellcolor[gray]{0.9}&\cellcolor[gray]{0.9}&Number of iterations      & 600 & 500     \\
		\multirow{-3}{*}{\rotatebox{90}{\cellcolor[gray]{0.9} lbm}}&\multirow{-3}{*}{\rotatebox{90}{\cellcolor[gray]{0.9} 05}}&\multirow{-3}{*}{\rotatebox{90}{\cellcolor[gray]{0.9} C}}&\multirow{-3}{*}{\rotatebox{90}{\cellcolor[gray]{0.9} 9000}}&\multirow{-3}{*}{\rotatebox{90}{\cellcolor[gray]{0.9} {Barrier}}}&Seed for random number generator & 13948 &	13948\\	
		
        \midrule
		\cellcolor[gray]{0.9}&\cellcolor[gray]{0.9}&\cellcolor[gray]{0.9}&\cellcolor[gray]{0.9}&\cellcolor[gray]{0.9}&Initial seed for the random number generator  & 42   & 42   \\    
		\cellcolor[gray]{0.9}&\cellcolor[gray]{0.9}&\cellcolor[gray]{0.9}&\cellcolor[gray]{0.9}&\cellcolor[gray]{0.9}&Number of simulated time steps  & 200 &  400  \\    
		\multirow{-3}{*}{\rotatebox{90}{\cellcolor[gray]{0.9} soma}}&\multirow{-3}{*}{\rotatebox{90}{\cellcolor[gray]{0.9} 13}}&\multirow{-3}{*}{\rotatebox{90}{\cellcolor[gray]{0.9} C}}&\multirow{-3}{*}{\rotatebox{90}{\cellcolor[gray]{0.9} 9500}}&\multirow{-3}{*}{\rotatebox{90}{\cellcolor[gray]{0.9} {Allreduce}}}&Number of simulated polymers      & 14000000 & 25000000     \\

		\midrule
		\cellcolor[gray]{0.9}&\cellcolor[gray]{0.9}&\cellcolor[gray]{0.9}&\cellcolor[gray]{0.9}&\cellcolor[gray]{0.9}&Density in five different states& \{100,0.1,0.1,0.1,0.1\}, & \{100,0.1,0.1,0.1,0.1\}   \\
  		\cellcolor[gray]{0.9}&\cellcolor[gray]{0.9}&\cellcolor[gray]{0.9}&\cellcolor[gray]{0.9}&\cellcolor[gray]{0.9}& Energy in five different states & \{0.0001,25,0.1,0.1,0.1\} & \{0.0001,25,0.1,0.1,0.1\}   \\  
		\cellcolor[gray]{0.9}&\cellcolor[gray]{0.9}&\cellcolor[gray]{0.9}&\cellcolor[gray]{0.9}&\cellcolor[gray]{0.9}&Size of the computational domain \{min,max\} & \{0,10\} &  \{0,10\} \\    
		\cellcolor[gray]{0.9}&\cellcolor[gray]{0.9}&\cellcolor[gray]{0.9}&\cellcolor[gray]{0.9}&\cellcolor[gray]{0.9}&Cell count for \{X,Y\}-direction       & \{8192,8192\} &  \{16384,16384\}  \\
  		\cellcolor[gray]{0.9}&\cellcolor[gray]{0.9}&\cellcolor[gray]{0.9}&\cellcolor[gray]{0.9}&\cellcolor[gray]{0.9}& Method to solve the linear system & Conjugate Gradient &   Conjugate Gradient \\    
		\cellcolor[gray]{0.9}&\cellcolor[gray]{0.9}&\cellcolor[gray]{0.9}&\cellcolor[gray]{0.9}&\cellcolor[gray]{0.9}& Solver convergence threshold using residual's least squares &  $1.0e^{-15}$ &   $1.0e^{-15}$  \\  
		\cellcolor[gray]{0.9}&\cellcolor[gray]{0.9}&\cellcolor[gray]{0.9}&\cellcolor[gray]{0.9}&\cellcolor[gray]{0.9}&Upper iterations limit for the linear solver in a step & 5000 & 5000\\	
    	\cellcolor[gray]{0.9}&\cellcolor[gray]{0.9}&\cellcolor[gray]{0.9}&\cellcolor[gray]{0.9}&\cellcolor[gray]{0.9}& Initial time-step & 0.004 &	0.004\\	
  		\cellcolor[gray]{0.9}&\cellcolor[gray]{0.9}&\cellcolor[gray]{0.9}&\cellcolor[gray]{0.9}&\cellcolor[gray]{0.9}& Simulation end times \{end time, end step\} & \{5, 100\} &	\{15, 100\}\\	
        \cellcolor[gray]{0.9}&\cellcolor[gray]{0.9}&\cellcolor[gray]{0.9}&\cellcolor[gray]{0.9}&\cellcolor[gray]{0.9}&Number of inner steps when using PPCG solver& 350 &	350\\
        \multirow{-10}{*}{\rotatebox{90}{\cellcolor[gray]{0.9} tealeaf}}&\multirow{-10}{*}{\rotatebox{90}{\cellcolor[gray]{0.9} 18}}&\multirow{-10}{*}{\rotatebox{90}{\cellcolor[gray]{0.9} C}}&\multirow{-10}{*}{\rotatebox{90}{\cellcolor[gray]{0.9} 5400}}&\multirow{-10}{*}{\rotatebox{90}{\cellcolor[gray]{0.9} {Allreduce}}}&Number of CG iterations before the Chebyshev method\mbox{$^\ddagger$} & 20 &	20\\

		\midrule
		\cellcolor[gray]{0.9}&\cellcolor[gray]{0.9}&\cellcolor[gray]{0.9}&\cellcolor[gray]{0.9}&\cellcolor[gray]{0.9}& \{density, energy\} in two ideal gas states & \{0.2,1\},\{1,2.5\}  & \{0.2,1\},\{1,2.5\}   \\    
		\cellcolor[gray]{0.9}&\cellcolor[gray]{0.9}&\cellcolor[gray]{0.9}&\cellcolor[gray]{0.9}&\cellcolor[gray]{0.9}&Logical mesh size for \{X,Y\}-direction  &  \{15360,15360\}  & \{61440,30720\}  \\    
		\cellcolor[gray]{0.9}&\cellcolor[gray]{0.9}&\cellcolor[gray]{0.9}&\cellcolor[gray]{0.9}&\cellcolor[gray]{0.9}& Physical mesh size for \{X,Y\}-direction \{Xmin,Ymin,Xmax,Ymax\} & \{0,0,10,10\}  & \{0,0,10,10\}   \\  
		\cellcolor[gray]{0.9}&\cellcolor[gray]{0.9}&\cellcolor[gray]{0.9}&\cellcolor[gray]{0.9}&\cellcolor[gray]{0.9}&Timestep frequency \{initial, rise, max\} timestep & \{0.04, 1.5, 0.04\} &   \{0.04, 1.5, 0.04\} \\    
		\multirow{-5}{*}{\rotatebox{90}{\cellcolor[gray]{0.9} clvleaf}}&\multirow{-5}{*}{\rotatebox{90}{\cellcolor[gray]{0.9} 19}}&\multirow{-5}{*}{\rotatebox{90}{\cellcolor[gray]{0.9} Fortran}}&\multirow{-5}{*}{\rotatebox{90}{\cellcolor[gray]{0.9} 12500}}&\multirow{-5}{*}{\rotatebox{90}{\cellcolor[gray]{0.9} {Allreduce}}}&Simulation end times \{end time, end step\} & \{0.5, 400\} &	\{0.5, 500\}\\	

		\midrule
		\cellcolor[gray]{0.9}&\cellcolor[gray]{0.9}&\cellcolor[gray]{0.9}&\cellcolor[gray]{0.9}&\cellcolor[gray]{0.9}&Number of sweep iterations  & 40   & 80   \\    
		\cellcolor[gray]{0.9}&\cellcolor[gray]{0.9}&\cellcolor[gray]{0.9}&\cellcolor[gray]{0.9}&\cellcolor[gray]{0.9}&Global number of grid cells along the \{X,Y,Z\}-dimension  & \{96,64,64\} &  \{128,64,64\} \\    
		\cellcolor[gray]{0.9}&\cellcolor[gray]{0.9}&\cellcolor[gray]{0.9}&\cellcolor[gray]{0.9}&\cellcolor[gray]{0.9}&Total number of energy groups  & 64 &   64 \\    
		\cellcolor[gray]{0.9}&\cellcolor[gray]{0.9}&\cellcolor[gray]{0.9}&\cellcolor[gray]{0.9}&\cellcolor[gray]{0.9}&Number of angles for each octant direction  & 32 &   32 \\    
		\multirow{-5}{*}{\rotatebox{90}{\cellcolor[gray]{0.9} minisweep}}&\multirow{-5}{*}{\rotatebox{90}{\cellcolor[gray]{0.9} 21}}&\multirow{-5}{*}{\rotatebox{90}{\cellcolor[gray]{0.9} C}}&\multirow{-5}{*}{\rotatebox{90}{\cellcolor[gray]{0.9} 17500}}&\multirow{-5}{*}{\rotatebox{90}{\cellcolor[gray]{0.9} {--}}}&Number of sweep blocks used to tile the Z-dimension & 8 &	8\\	
  
		\midrule
		\cellcolor[gray]{0.9}&\cellcolor[gray]{0.9}&\cellcolor[gray]{0.9}&\cellcolor[gray]{0.9}&\cellcolor[gray]{0.9}&Number of nr  & 173  & 325   \\    
		\cellcolor[gray]{0.9}&\cellcolor[gray]{0.9}&\cellcolor[gray]{0.9}&\cellcolor[gray]{0.9}&\cellcolor[gray]{0.9}&Number of nt  & 361 &  450  \\    
		\multirow{-3}{*}{\rotatebox{90}{\cellcolor[gray]{0.9} pot3d}}&\multirow{-3}{*}{\rotatebox{90}{\cellcolor[gray]{0.9} 28}}&\multirow{-3}{*}{\rotatebox{90}{\cellcolor[gray]{0.9} Fortran}}&\multirow{-3}{*}{\rotatebox{90}{\cellcolor[gray]{0.9} 495000\mbox{$^\amalg$}}}&\multirow{-3}{*}{\rotatebox{90}{\cellcolor[gray]{0.9} {Allreduce}}}& Number of np  & 1171 &	2050\\	
  
        \midrule
		\cellcolor[gray]{0.9}&\cellcolor[gray]{0.9}&\cellcolor[gray]{0.9}&\cellcolor[gray]{0.9}&\cellcolor[gray]{0.9}&Number of particles to the cube  & $210^3$   & $350^3$   \\    
		\cellcolor[gray]{0.9}&\cellcolor[gray]{0.9}&\cellcolor[gray]{0.9}&\cellcolor[gray]{0.9}&\cellcolor[gray]{0.9}&Number of time-steps  & 80 &  100  \\  		\multirow{-3}{*}{\rotatebox{90}{\cellcolor[gray]{0.9} sph-exa}}&\multirow{-3}{*}{\rotatebox{90}{\cellcolor[gray]{0.9} 32}}&\multirow{-3}{*}{\rotatebox{90}{\cellcolor[gray]{0.9} C++14}}&\multirow{-3}{*}{\rotatebox{90}{\cellcolor[gray]{0.9} 3400}}&\multirow{-3}{*}{\rotatebox{90}{\cellcolor[gray]{0.9} {Allreduce}}}&How often output file shall be written\mbox{$^\mathparagraph$} & -1 &	-1\\
  
		\midrule
		\cellcolor[gray]{0.9}&\cellcolor[gray]{0.9}&\cellcolor[gray]{0.9}&\cellcolor[gray]{0.9}&\cellcolor[gray]{0.9}&Log to base 2 of the box dimension\mbox{$^\bigstar$}  & 5  & 5   \\    
		\cellcolor[gray]{0.9}&\cellcolor[gray]{0.9}&\cellcolor[gray]{0.9}&\cellcolor[gray]{0.9}&\cellcolor[gray]{0.9}&Log to base 2 of the grid dimension\mbox{$^\ast$}  & 9 &  10  \\    
		\multirow{-3}{*}{\rotatebox{90}{\cellcolor[gray]{0.9} hpgmgfv}}&\multirow{-3}{*}{\rotatebox{90}{\cellcolor[gray]{0.9} 34}}&\multirow{-3}{*}{\rotatebox{90}{\cellcolor[gray]{0.9} C}}&\multirow{-3}{*}{\rotatebox{90}{\cellcolor[gray]{0.9} 16700}}&\multirow{-3}{*}{\rotatebox{90}{\cellcolor[gray]{0.9} {Allreduce}}}& Number of time-steps & 300 &	300\\	

        \midrule  
		\cellcolor[gray]{0.9}&\cellcolor[gray]{0.9}&\cellcolor[gray]{0.9}&\cellcolor[gray]{0.9}&\cellcolor[gray]{0.9}&Global X-dimension size  & 24000 10000 &  192000 10000 \\    
		\cellcolor[gray]{0.9}&\cellcolor[gray]{0.9}&\cellcolor[gray]{0.9}&\cellcolor[gray]{0.9}&\cellcolor[gray]{0.9}&Global Z-dimension size      & 3000 1250 &   24000  1250  \\
		\cellcolor[gray]{0.9}&\cellcolor[gray]{0.9}&\cellcolor[gray]{0.9}&\cellcolor[gray]{0.9}&\cellcolor[gray]{0.9}&Number of time-steps  & 600  & 600   \\    
		\cellcolor[gray]{0.9}&\cellcolor[gray]{0.9}&\cellcolor[gray]{0.9}&\cellcolor[gray]{0.9}&\cellcolor[gray]{0.9}&Output over N number of time-steps  & 100 &   100 \\    
		\multirow{-5}{*}{\rotatebox{90}{\cellcolor[gray]{0.9} weather}}&\multirow{-6}{*}{\rotatebox{90}{\cellcolor[gray]{0.9} 35}}&\multirow{-6}{*}{\rotatebox{90}{\cellcolor[gray]{0.9} Fortran}}&\multirow{-6}{*}{\rotatebox{90}{\cellcolor[gray]{0.9} 1100}}&\multirow{-6}{*}{\rotatebox{90}{\cellcolor[gray]{0.9} {--}}}&Model number to use\mbox{$^\mathsection$} & 6 &	6\\	
		\bottomrule
	\end{tabular}
	\begin{tablenotes}
        \item \mbox{$^\ddagger$} Starting the Chebyshev method requires providing approximations of the minimum and maximum eigenvalues.\\
        \item \mbox{$^\amalg$} This includes the Line of Codes (LOC) from HDF5 library as well.\\
        \item \mbox{$^\mathparagraph$} The automatic generation of the input conditions for all provided particles has been added to the source code for testing purposes.\\
        \item \mbox{$^\bigstar$} The finest grid comprises boxes of size $32^3$ grid points and 
        \mbox{$^\ast$} a total of $512^3$ (tiny suite) and $1024^3$ (medium suite) grid points.\\
        \item \mbox{$^\mathsection$} Models: (1) Colliding Thermals, (2) Rising Thermals, (3) Mountain Gravity Waves, (4) Turbulence, (5) Density Current, (6) Injection
        
        \noindent\rule[0.5ex]{\linewidth}{1pt}
    \end{tablenotes}
    \end{threeparttable}
    \end{adjustbox}
    \vspace{-2em}
\end{table}

%% file: figures/SPEC_brief.tex
	\begin{table}[t]
		\centering
		\caption{Numeric and domain data of SPEChpc 2021 suite.}
        \label{tab:SPECbrief}
		\begin{adjustbox}{width=0.48\textwidth}
				\setlength\extrarowheight{-0.7pt}
                \setlength\tabcolsep{2pt}
                \begin{tabular}[fragile]{c>{~}cy}
                	\toprule
                    \rowcolor[gray]{0.9}
                	\cellcolor[gray]{0.9} Name& Numerical brief information  &  Application domain\\
                	\midrule
                	\cellcolor[gray]{0.9}\multirow{-1}{*}{\rotatebox{0}{\cellcolor[gray]{0.9} \CODE{lbm}}}&Lattice-Boltzmann Method D2Q37   & 2D CFD solver  \\    
                	\cellcolor[gray]{0.9}\multirow{-1}{*}{\rotatebox{0}{\cellcolor[gray]{0.9} \CODE{soma}}}&Monte-Carlo acceleration for soft coarse grained polymers       & Physics or polymeric systems  \\
                	\cellcolor[gray]{0.9}\multirow{-1}{*}{\rotatebox{0}{\cellcolor[gray]{0.9} \CODE{tealeaf}}}&Solving the linear heat conduction equation on a 2D regular grid    & Physics or high energy physics  \\
                    \cellcolor[gray]{0.9}&using a 5-point stencil with implicit solvers   &   \\
                    \cellcolor[gray]{0.9}\multirow{-1}{*}{\rotatebox{0}{\cellcolor[gray]{0.9} \CODE{cloverleaf}}}&Solving compressible Euler equations on a 2D Cartesian grid   & Physics or high energy physics  \\
                    \cellcolor[gray]{0.9}&using an explicit second-order accurate method   &   \\
                    \cellcolor[gray]{0.9}\multirow{-1}{*}{\rotatebox{0}{\cellcolor[gray]{0.9}  \CODE{minisweep}}}&A successor to the well-known Sweep3D benchmark   & Radiation transport in nuclear engineering  \\
                    \cellcolor[gray]{0.9}\multirow{-1}{*}{\rotatebox{0}{\cellcolor[gray]{0.9} \CODE{pot3D}}}&Computing potential field solutions using a preconditioned CG    & Solar physics  \\
                    \cellcolor[gray]{0.9}& sparse solver for the Laplace equation in 3D spherical coordinates   &   \\
                    \cellcolor[gray]{0.9}\multirow{-1}{*}{\rotatebox{0}{\cellcolor[gray]{0.9} \CODE{sph-exa}}}&Smoothed Particle Hydrodynamics, a meshless Lagrangian method   & Astrophysics and cosmology  \\
                    \cellcolor[gray]{0.9}\multirow{-1}{*}{\rotatebox{0}{\cellcolor[gray]{0.9} \CODE{hpgmgfv}}}&Finite-volume-based High Performance Geometric Multigrid   & Cosmology, astrophysics, combustion  \\
                    \cellcolor[gray]{0.9}&solving variable-coefficient elliptic problems on Cartesian grids   &   \\
                	\cellcolor[gray]{0.9}\multirow{-1}{*}{\rotatebox{0}{\cellcolor[gray]{0.9} \CODE{weather}}}& A traditional finite-volume control flow   & Atmospheric weather and climate  \\    
                	\bottomrule
                \end{tabular}
		\end{adjustbox}
    \vspace{-1.2em}
	\end{table}

%% file: figures/tab_systems.tex
	\begin{table}[t]
		\centering
		\caption{Key hardware and software attributes of systems.} 
		\label{tab:systems}
		\begin{adjustbox}{width=0.48\textwidth}
				\setlength\extrarowheight{-0.7pt}
                \setlength\tabcolsep{2pt}
                \Huge
                \begin{tabular}[fragile]{c>{~}lvx}
                	\toprule
                	\rowcolor[gray]{0.9}
                	\cellcolor[gray]{0.9}&Systems  &  ClusterA & ClusterB \\
                	\midrule
                	\cellcolor[gray]{0.9}&Intel Processor   & Xeon Ice Lake  & Xeon Sapphire Rapids \\    
                	\cellcolor[gray]{0.9}&Processor Model       & Platinum 8360Y  & Platinum 8470   \\
                	\cellcolor[gray]{0.9}&Base clock speed  & $2.4$~\GHZ\  &  $2.0$~\GHZ\ \\
                	\cellcolor[gray]{0.9}&Physical cores per node     & 72    & 104      \\
                	\cellcolor[gray]{0.9}&ccNUMA domains per node   & 4  & 8   \\
                	\cellcolor[gray]{0.9}&Sockets per node   & 2  & 2   \\
                	\cellcolor[gray]{0.9}&Per-core L1/L2 cache &  $48$~\KiB\ (L1) + $1.25$~\MiB\ (L2)  & $48$~\KiB\ (L1) + $2$~\MiB\ (L2) \\
                	\cellcolor[gray]{0.9}&Shared LLC &  $54$~\MiB\ (L3)  & $105$~\MiB\ (L3) \\
                	\cellcolor[gray]{0.9}&Memory per node & $4\times 64$~\GiB\  & $8\times 128$~\GiB\ \\
                	\cellcolor[gray]{0.9}&Socket memory type & 8 channels DDR4-3200  & 8 channels DDR5-4800 \\
                	\cellcolor[gray]{0.9}&Theor. socket memory bandwidth &  $2\times 102.4$~\GBS\ & $4\times 76.8$~\GBS\ \\
                    \multirow{-10}{*}{\rotatebox{90}{\cellcolor[gray]{0.9} Micro-architecture}}&Thermal design power & 250 W & 350 W\\                	
                    \midrule
                	\cellcolor[gray]{0.9}&Node interconnect     & HDR100 Infiniband  & HDR100 Infiniband    \\
                	\cellcolor[gray]{0.9}&Interconnect topology  & Fat-tree  & Fat-tree \\
                	\cellcolor[gray]{0.9}&Raw bandwidth per link \& direction &  $100$~\GBPS\ & $100$~\GBPS\   \\
                    \cellcolor[gray]{0.9}&Parallel filesystem (capacity) & Lustre-based ($3.5$~PB) & Lustre-based ($3.5$~PB)\\ 
                    \multirow{-5}{*}{\rotatebox{90}{\cellcolor[gray]{0.9} Network}}&Aggregated parallel I/O bandwidth & $>20$~\GBS\ & $>20$~\GBS\ \\
                	
                	\midrule
                	\cellcolor[gray]{0.9}&Compiler    & Intel v2022u1   & Intel v2022u1 \\
                	\cellcolor[gray]{0.9}&Optimization flags & -O3 -qopt-zmm-usage=high & -O3 -qopt-zmm-usage=high \\
                	\cellcolor[gray]{0.9}&SIMD &-xCORE-AVX512 &  -xCORE-AVX512\\
                	\cellcolor[gray]{0.9}&Message passing library  & Intel \verb.MPI. v2021u7    &  Intel \verb.MPI. v2021u7 \\
                	\multirow{-5}{*}{\rotatebox{90}{\cellcolor[gray]{0.9} Software}}&Operating system    &  AlmaLinux v8.8  &  AlmaLinux v8.8 \\
                	
                	\midrule
                   \cellcolor[gray]{0.9}&\CODE{ClusterCockpit}    &  v1u0.0  &  v1u0.0  \\           
                	\cellcolor[gray]{0.9}&\CODE{ITAC} version      & v2021u6  & v2021u6 \\
                	\cellcolor[gray]{0.9}&\CODE{ITAC} flags      & -trace -tcollect  & -trace -tcollect \\
                	\cellcolor[gray]{0.9}&\CODE{LIKWID} version  & 5.2.2  & 5.2.2/saprap1 (beta)\\
                	\multirow{-5}{*}{\rotatebox{90}{\cellcolor[gray]{0.9} Tools}}&\CODE{LIKWID} flags  & -g MEM\_DP/L3/L2    & -g MEM\_DP/L3/L2 \\
                	\bottomrule
                \end{tabular}
		\end{adjustbox}
    \vspace{-2em}
	\end{table}